\documentclass[fleqn,usenatbib]{mnras}

\usepackage[T1]{fontenc}
\usepackage{ae,aecompl}

\usepackage{graphicx}	
\usepackage{amsmath}	
\usepackage{amssymb}	
\usepackage{newtxtext,newtxmath}

\newcommand{\Mh}{$h^{-1}$M$_\odot$}

\usepackage{relsize}

\newcommand{\eoh}{EAGLE100}

\newcommand{\Lgr}{\larger[2]}
\newcommand{\Smr}{\smaller[2]}
\newcommand{\lgr}{\larger[1.4]}
\newcommand{\smr}{\smaller[1.4]}

\newcommand{\galform}{\smr{GALFORM}\lgr{}}

\newcommand{\dhalo}{Dhalo}

\newcommand{\hbt}{\smr{HBT}\lgr}

\newcommand{\rockstart}{\smr{ROCKSTAR}\lgr{}-\smr{T}\lgr{}\Smr{REE}\Lgr{}}
\newcommand{\rockstar}{\smr{ROCKSTAR}\lgr{}}
\newcommand{\constenttrees}{\smr{C}\lgr{}\Smr{ONSISTENT}\Lgr{}\smr{T}\lgr{}\Smr{REES}\Lgr}

\newcommand{\subfindt}{\smr{SUBFIND}\lgr{}-\smr{T}\lgr{}\Smr{REE}\Lgr{}}
\newcommand{\subfind}{\smr{SUBFIND}\lgr{}}
\newcommand{\dtrees}{\smr{D-TREES}\lgr{}}

\newcommand{\velociraptort}{\smr{VELOCI}\lgr{}\Smr{RAPTOR}\Lgr{}-\smr{T}\lgr{}\Smr{REE}\Lgr{}}
\newcommand{\velociraptor}{\smr{VELOCI}\lgr{}\Smr{RAPTOR}\Lgr{}}
\newcommand{\treefrog}{\smr{T}\lgr{}\Smr{REE}\Lgr{}\smr{F}\lgr{}\Smr{FROG}\Lgr{}}

\newcommand{\mstellar}{M$_{\mathrm{Stellar}}$}
\newcommand{\mhotgas}{M$_{\mathrm{Hot\, Gas}}$}
\newcommand{\mcoldgas}{M$_{\mathrm{Cold\, Gas}}$}
\newcommand{\inststr}{instantaneous SFR}

\title[Halo Merger Tree Comparison]{Halo Merger Tree Comparison: Impact on Galaxy Formation Models}

\author[Jonathan S. G\'omez et al.]{
Jonathan S. G\'omez,$^{1, 2}$\thanks{E-mail: jsgomez1@uc.cl}
N. D. Padilla,$^{1}$
J. C. Helly,$^{3}$
C. G. Lacey,$^{3}$
C. M. Baugh,$^{3}$ 
C. D. P. Lagos$^{4}$
\\
$^{1}$Instituto de Astrof\'\i sica, Pontificia Universidad Cat\'olica de Chile, Av. Vicu\~na Mackenna 4860, Stgo., Chile.\\
$^{2}$Departamento de F\'isica Te\'orica, Universidad Aut\'onoma de Madrid, E-28049 Cantoblanco, Madrid, Spain.\\
$^{3}$Institute for Computational Cosmology, Department of Physics, University of Durham, South Road, Durham DH1 3LE, UK.\\
$^{4}$International Centre for Radio Astronomy Research (ICRAR), M468, University of Western Australia, 35 Stirling Hwy, Crawley, WA 6009, Australia.
}

\date{Accepted XXX. Received YYY; in original form ZZZ}

\pubyear{2021}

\begin{document}
\label{firstpage}
\pagerange{\pageref{firstpage}--\pageref{lastpage}}
\maketitle

\begin{abstract}
We examine the effect of using different halo finders and merger tree building algorithms on galaxy properties predicted using the \galform~semi-analytical model run on a high resolution, large volume dark matter  simulation. The  halo finders/tree builders \hbt, \rockstar, \subfind~and \velociraptor~differ in their definitions of halo mass, on whether only spatial or phase-space information is used, and in how they distinguish satellite and main haloes; all of these features have some impact on the model~galaxies, even after the trees are post-processed and homogenised by \galform.   The stellar mass function is insensitive to the halo and merger tree finder adopted.  However, we find that the number of central and satellite galaxies in \galform~does depend  {slightly} on the halo finder/tree builder. 
The number of galaxies without resolved subhaloes depends strongly on the tree builder, with \velociraptor, a phase-space finder, showing the largest population of such galaxies. 
 {The distributions of stellar masses, cold and hot gas masses, and star formation rates} agree well between different halo finders/tree builders. However, because \velociraptor~{has} more early progenitor haloes, with these trees \galform~produces slightly higher star formation rate densities at high redshift,  smaller galaxy sizes, {and larger stellar masses for the spheroid component}. Since in all cases these differences are small {we conclude} that,  when all of the trees are processed  {so that the main progenitor mass increases monotonically}, the predicted \galform~galaxy populations are stable and consistent for these four halo finders/tree builders.
\end{abstract}


\section{Introduction}

In the Lambda Cold Dark Matter ($\Lambda$CDM) model, galaxy formation and evolution are  directly linked to the formation and evolution of dark matter haloes. Stars are formed within cold baryonic gas clouds resulting from the cooling of hot gas, which is heated by shocks as haloes of dark matter collapse gravitationally \citep{Binney1977, Rees1977, White1978}.


The formation and evolution of dark matter haloes in $\Lambda$CDM is well understood due to the simplicity of the physics  {-- {to a good approximation one can assume that dark matter} interacts only via gravity --} which is readily tackled using simulations. However, the evolution of the baryonic component is more uncertain and requires choices to be made regarding the subgrid physics (see the review by \citealt{Somerville:2015}). One of the leading approaches for modelling the formation and evolution of galaxies in  $\Lambda$CDM is semi-analytical modelling (SAM; see  e.g. \citealt{Cole1991}, \citealt{Lacey1991} and \citealt{White1991} for the first examples of such models).  
This approach uses the evolution of dark matter haloes as obtained from   Monte-Carlo prescriptions \citep{Kauffmann19931,Kauffmann19932,Cole1994} or N-body simulations \citep{Somerville2008,Benson2012,Lacey2016} and couples this to simplified physical models of the baryonic physics governing galaxy formation (for reviews see \citealt{Baugh2006} and \citealt{Benson2010b}).

In a cosmological N-body simulation the mass resolution sets the minimum halo mass that can be reliably detected. Haloes grow by mergers or via smoothly accreting material. The merging process does not immediately produce a relaxed smooth halo; the remnants of earlier generations of haloes are often detectable as self-bound substructures (subhaloes or satellites) within the new halo. \cite{Knebe2011, Knebe2013} demonstrated that most widely-used halo finder codes generate similar halo properties  {since most start with a standard percolation algorithm. Consequently they usually obtain similar} halo and subhalo mass functions. However, poorly  resolved halos or dense environments can be problematic for identifying substructures  for some halo finders (or substructure finders; \citealt{Muldrew2011, Onions2013, Elahi2013}). Some finders are able to identify arbitrary levels of nested satellites within satellites and also identify the background mass distribution in a halo as the main subhalo. Thus, it is important to distinguish between primary or main haloes and the satellite subhaloes that they contain which are the remnants of earlier generations of accreted haloes.

Therefore, in order to fully understand how the choice of halo finder/tree builder affects how a particular SAM models galaxy evolution, one can use the output of a single dark matter only simulation  {processed through} different methods.   {The resulting}  haloes, subhaloes, merger trees and galaxy catalogues 
 { can be analysed to determine differences between the algorithms}. 

There are many halo finder codes, going back to the  spherical-overdensity (SO) method first mentioned by \cite{Press1974} and the Friends-of-Friends (FoF) algorithm introduced by \citet{Davis1985}. Many codes have also been designed to build merger trees. This is due to the need for more efficient algorithms  as simulations follow ever more particles and improve their resolution, and also due to new approaches followed to detect main and satellite haloes and to connect them into merger trees. 

Although there have been several halo finder and merger tree builder comparisons \citep{Knebe2011, Onions2012, Srisawat2013, Lee2014}, a study of their effect on a single SAM has not been carried out to date. Several authors have compared the outputs of different SAMs after applying them to a single dark-matter only simulation (see for instance \citealt{DeLucia:2010,Lee2014, knebe2015, Knebe2017a, Knebe2017b, Pujol2017, Asquith2018, Cui2018, Favole2020}). {In these works} the dark-matter only simulation {was} analysed with a single halo finder and merger tree builder, and all SAMs {were} run using {the same} halo {and} tree catalogues.  In several cases, the SAMs involved were designed to be run on a different halo and tree finder {from that used in the comparison}. These studies usually {focused} on the differences between the galaxy outputs of the different SAMs, without studying the differences arising from using a different halo finder and tree builder with respect to the ones usually used in each SAM.

Different SAM codes not only use different treatments of  baryonic physics, they are also sensitive to the way {in which} the merger trees  are constructed. Hence, it is important to understand the effects of the latter to be able to isolate the differences between SAMs that come from their particular treatment of subgrid physics.

We examine the effect of using different halo finder/merger tree building algorithms on the galaxy properties predicted using a single SAM, namely the \galform~ SAM coupled to a cosmological N-body simulation \citep{Cole2000,Lacey2016}. We use the implementation {of \galform~in the Planck cosmology} by \citet{Baugh2019}.  {This SAM features an  algorithm called \dhalo~which ensures the monotonicity of the growth of main halo masses, which in turn can act to homogenize the outputs of different finders.}  This comparison is not intended to decide whether one specific merger tree finder is better or worse than the others, but to study their effects on the predicted galaxy population. We propose this step as a necessary ingredient in future SAM comparison projects, to allow {one} to isolate the factors contributing to the differences between the outputs of different models of galaxy formation.

The outline of this paper is as follows. In Section \ref{merger_tree_builder_descrip}, we introduce the details of the dark matter N-body simulations used to construct the merger trees that are fed into \galform~to construct the galaxy population, and we overview the halo merger tree building algorithms \hbt~\citep{Han2012}, \rockstar~\citep{Behroozi2013a}, \subfind~\citep{Springel2001} and \velociraptor~\citep{Elahi2011} {compared} in this work. In Section \ref{SAM description}, we describe the processing of  merger trees by the \dhalo~ algorithm, and the version of \galform~used in this study. Then, in Section \ref{section: Merger Trees} we compare the {halo properties output by} different merger trees. We demonstrate the impact of using different halo merger trees on  \galform~galaxies in  Section \ref{section: gal_galform}.  In Section \ref{conclusions}, we summarize and present our conclusions.

\section{Dark Matter only simulation and halo and tree finders}
\label{merger_tree_builder_descrip}

We use one of the dark matter only simulations from the EAGLE simulation suite \citep{Schaye2015}, to study  the impact of different merger tree builders on a SAM of galaxy formation. This simulation adopts the \citet{Planck-Collaboration2014} cosmological parameters, shown in Table \ref{tab_parameters} along with other key parameters. This simulation calculates the evolution of dark matter in a periodic volume with comoving size $100$ Mpc on a side (hereafter referred as \eoh) and a dark matter particle mass $m_{\mathrm{DM}}=6.57\times10^6 h^{-1}M_{\odot}$.

Our study makes use of several combinations of halo finders and tree builders as listed in Table \ref{table:halo finders}. Each tree builder is designed to work with a particular halo finder, with the goal of determining what are their effects on a SAM of galaxy formation. The process of generating halo merger trees suitable for use with \galform~ consists of three steps, which make use of different algorithms.  The terminology we adopt in this study for the self-bound objects is as follows:

    \begin{itemize}
    \item Halo  {and subhalo} finder: Produces a catalogue of self-bound dark matter structures in approximate dynamical equilibrium  {also referred to as haloes.  Then an algorithm that searches for substructures or overdensities within these haloes is run.  Its output is then} processed to produce \textit{subhalos} that are classified as \textit{main subhalos} or \textit{satellite subhalos}. This is done for each simulation snapshot.  
    \item Tree Builder: Identifies progenitors and descendants for each \textit{subhalo} for all snapshots.
    \item \dhalo: The subhalos are processed into \textit{central subhalos} and \textit{satellite subhalos} by \textit{Dhalos}, applying the algorithm described in Section~\ref{Constructing Dhalo Merger Trees} (see also Appendix A3 of \citet{Jiang2014}).
    \end{itemize}

The specific definition of halos in \galform~ is provided by the \dhalo~algorithm. Dhalos are the largest gravitationally bound structures in the dark matter that are in approximate dynamical equilibrium, which by definition are not contained within any similar larger structure.  They may be referred to with different terminology in the different halo finder codes, but we will {follow the \galform~terminology and} use \dhalo~for the whole gravitationally collapsed structure, and central subhalo for the most prominent subhalo in the Dhalo. Here we refer to each combination of halo finder and tree builder as a `subhalo merger tree builder'. The combinations {that} we use in this paper are presented in Table \ref{table:halo finders}.

    \begin{table*}
    \normalsize
    \caption{Cosmological and numerical parameters of the N-body simulation used in this work.}
    \begin{tabular}{lcccc}
    \hline\hline
    Parameter        & \multicolumn{2}{c|}{Meaning}                                          & Value   & Reference                       \\
    \hline
    \hline
    $\Omega_m$       & \multicolumn{2}{c|}{Matter density parameter}      & 0.307   &                                 \\
    $\Omega_\Lambda$ & \multicolumn{2}{c|}{{Vacuum} energy density parameter} & 0.693   &                                 \\
    $h$              & \multicolumn{2}{c|}{$H_0/(100$ km s$^{-1}$ Mpc$^{-1})$}                        & 0.6777  & \cite{Planck-Collaboration2014} \\
    $n_s$            & \multicolumn{2}{c|}{Primordial power spectrum index}                           & 0.9611  &                                 \\
    $\sigma_8$       & \multicolumn{2}{c|}{rms linear density fluctuations in spheres of radius} $8 h^{-1}$Mpc & 0.8288  &                                 \\
    \hline
    \hline
    Simulation   &	$L_{\mathrm{box}}$/Mpc & $N_{\mathrm{p}}$ & $m_{\mathrm{dmp}} \,h^{-1}M_{\odot}$   & Reference         \\
    \hline
    \hline
    \eoh         &  100                    & $1504^3$         & $6.57 \times 10^6$       & \cite{Schaye2015} \\
    \hline
    \end{tabular}
    \label{tab_parameters}
    \end{table*}

    \begin{table}
    \caption{Subhalo merger tree builders used. The first column gives the combined names of the halo finders and tree builders, column (2) gives the halo finder names and column (3) gives the tree builder names.}
    \begin{tabular}{ccc}
    \hline
    \hline
    Combined name  & halo finder           &  tree builder         \\
    \hline
    \hline
    \hbt           & Halo finding and tree &\hskip -.65cm building in same process \\
                   \citep{Han2012}                                 \\
    \hline
    \rockstart     & \rockstar             & \constenttrees        \\
                   & \citep{Behroozi2013a} & \citep{Behroozi2013b} \\
    \hline
    \subfindt      & \subfind              & \dtrees               \\
                   & \citep{Springel2001}  & \citep{Jiang2014}     \\
    \hline
    \velociraptort & \velociraptor         & \dtrees               \\ 
                   & \citep{Elahi2011}     & \citep{Jiang2014}     \\ 
    \hline
    \end{tabular}
    \label{table:halo finders}
    \end{table}

The four merger tree builder combinations were applied to the \eoh~ dark matter only simulation, generating merger trees using a total of 201 snapshots (sn), from sn=0 to sn=200, distributed 
 between $z=20$ and $z=0$, with the exception of \velociraptor~{for which we used} a total of 200 
 snapshots.

All  halo finders considered use, as a first step, the standard Friends-of-Friends  algorithm (hereinafter FoF or 3DFoF). Some finders opt to supplement this with a more sophisticated search for haloes and their substructures, such as studying the particles in phase-space. We can split the halo finders into two categories, the ones that identify 3D overdensities using particle positions, and those that identify 6D overdensities in phase-space, using particle positions and velocities. For more global similarities and differences between the different halo merger tree builders see Table~\ref{table: halo finder parameters}. The rest of this section provides a short description of each algorithm.

    \begin{table*}
    \caption{Parameters for halo finders and tree builders.}
    \begin{tabular}{ccccc}
    \hline
    \hline
    Parameter                                    & \hbt      & \rockstart               & \subfindt                     & \velociraptort           \\
    \hline
    \hline
    Linking length for first step                & 0.2       & 0.28                     & 0.2                           & 0.2                      \\
    \hline
    Minimum number of DM particles for subhaloes & 20        & 2                        & 20                            & 20                       \\ 
    \hline
    First step for search                        & 3DFoF     & 3DFoF                    & 3DFoF                         & 3DFoF                    \\
    \hline
    Subsequent step mechanism for search         & 3DFoF     & {6DFoF}                  & {3D density field}            & 6DFoF                    \\
    \hline
    Information used by the subsequent step      & Positions & Positions and Velocities & Positions                     & Positions and Velocities \\
    \hline
    \end{tabular}
    \label{table: halo finder parameters}
    \end{table*}

\subsection{HBT} 
\label{subsubsection: HBTtree}

Hierarchical Bound Tracing\footnote{\href{https://github.com/Kambrian/HBTplus}{https://github.com/Kambrian/HBTplus}} (HBT; \citealt{Han2012}) is a tracking algorithm and halo finder that works in the time domain,  following structures from one timestep to the next. At every snapshot, isolated groups are identified with a standard FoF algorithm with the usual linking length of $b=0.2$ times the mean interparticle separation \citep{Davis1985}.  {Each group is then subject to an iterative unbinding procedure. Particles with positive total energy are removed until only bound particles remain. If the number of bound particles is above a minimum threshold, the candidate is recorded as a self-bound group.  This procedure is common to all finders used here.} For FoF groups with no progenitor, the self-bound part is identified as the start of a new merger tree branch. 
In other cases, the FoF group contains one or more self-bound subhaloes having progenitors identified in earlier snapshots.
Within each FoF group, the most massive subhalo is defined as the main subhalo which is the dominant subhalo within the host FoF group. This process returns exclusive\footnote{If particles are allowed to be members of only one subhalo, (i.e. particles in satellites are not included in the particle ID list of the main subhalo, and particles in overlapping subhaloes are assigned to just one of the two), then the subhaloes are said to be exclusive; otherwise they are inclusive.} arbitrarily shaped gravitationally bound objects which in our runs are set to contain at least 20 particles. Subhalo masses are simply the sum of the masses of their assigned dark matter particles.

Unlike other algorithms, \hbt~builds subhalo merger trees and finds the particle membership for every subhalo at every snapshot after its birth as part of a single process. Starting from the highest redshift, subhaloes are tracked down to later snapshots to link to their descendant subhaloes by generating a merger tree down to the main subhalo level. The particles contained within these main subhaloes are then followed explicitly through subsequent snapshots. To extend the merger tree down to the satellite subhalo level, \hbt~continues the tracing of merged branches, identifying the set of remaining self-bound particles for every progenitor subhalo. These self-bound remnants are defined as descendant subhaloes of their progenitors. When two or more subhaloes are linked to a common descendant subhalo, the algorithm compares the masses of the self-bound particles of the progenitor subhaloes, and defines their self-bound remnants, except the most massive remnant, as satellite subhaloes. As a result, every subhalo identified by \hbt~must have an explicit progenitor that traces back before infall, with no missing link along its evolution history. This means that any satellites in the first snapshot are not included as such in the trees. The current main subhalo is re-defined to be the self-bound part of all the particles in the FoF halo, excluding satellite particles allowing growth by smooth accretion, while its main progenitor is defined as the one that produced the most massive remnant. The tracking process is then continued for all the subhaloes, including main halos and satellites down to the final output of the simulation. As all subhaloes have at least one progenitor, all subhaloes have a descendant subhalo (except {at} the last snapshot). When a merger occurs, satellite subhaloes are tracked down to the lowest redshift snapshot; if its number of particles drops below 20,  { the satellite is assumed to have undergone a merger with the main subhalo and a record of the merged satellite is kept.}

\subsection{ROCKSTAR - CONSISTENTTREES} 

\subsubsection{Halo finder}
\label{subsubsection: rockstar}

\rockstar~\citep{Behroozi2013a} is a phase-space halo finder\footnote{\href{https://bitbucket.org/gfcstanford/rockstar}{https://bitbucket.org/gfcstanford/rockstar}} that attempts to maximize halo consistency across simulation snapshots \citep{Behroozi2013b}. \rockstar~first identifies FoF groups with a larger than usual linking length of $b=0.28$ times the mean interparticle separation, which links together the same particles as FoF with the standard linking length, plus additional ones, thereby  producing larger structures within which substructures are then found and post-processed. 

For each FoF group, particle positions and velocities are normalized by the group position and velocity dispersions such that for two particles $p_1$ and $p_2$ in a given group, they define a distance metric as
    \begin{equation}
    d(p_1,p_2) = \left[ \frac{(x_1 - x_2)^2}{\sigma_x^2} + \frac{(v_1 - v_2)^2}{\sigma_v^2} \right]^{1/2},
    \end{equation}
where $\sigma_x$ and $\sigma_v$ are the particle position and velocity dispersion, respectively, for the given FoF group. 

Within these groups \rockstar~builds a hierarchy of  subgroups using a phase-space linking length $d(p_1, p_2)$ that is progressively and adaptively chosen such that a constant fraction $f=0.7$ of the particles are linked together with at least one other particle in different levels of FoF subgroups as the process is repeated in each subgroup of the FoF group. The first and the most massive of the substructures found in this way corresponds to the main subhalo of the FoF group. Once the code finds no further substructure levels, \rockstar~converts FoF subgroups into subhaloes by exploring the different FoF subgroup levels starting from the deepest level and assigning particles to the closest subgroup in phase-space. Then the gravitational potentials of all particles are calculated using a modified Barnes \& Hut method \citep{Barnes1986} and this is used to unbind particles. Subhalo centres are defined by averaging particle positions at the FoF hierarchy level minimising Poisson error, which amounts in practice to averaging positions in a small region close to the phase-space density peak. The group masses adopted in this work for \rockstar~correspond to the sum of the masses of the particles listed as belonging to the groups. The particle membership list of a subhalo is exclusive and is made up of particles close in phase-space to the subhalo centre.\\

\subsubsection{Tree builder}

The \constenttrees~algorithm\footnote{\href{https://bitbucket.org/pbehroozi/consistent-trees}{https://bitbucket.org/pbehroozi/consistent-trees}} \citep{Behroozi2013b} first matches subhaloes between snapshots by identifying descendant subhaloes as those that contain the {largest number} of particles from the progenitor subhalo. It then cleans up this initial list taking into account the velocities and positions of progenitors and descendants, as well as their mass profiles. If a calculation suggests a missed satellite subhalo that would be too close to the centre of the larger subhalo {to be identified directly}, or spurious mass changes, these defects are repaired by substituting  {estimated} subhalo properties instead of the properties returned by the halo finder. Thus, the final masses produced by \rockstar~are not given by the sum of the masses of particles. A subhalo with no descendant is assumed to merge with the subhalo that exerts the strongest tidal field on it. If there is no such subhalo the progenitor is assumed to have been spurious and this branch is pruned from the merger tree. This process helps to ensure accurate mass accretion histories and merger rates for satellite and main subhaloes. If a satellite subhalo merges with a main subhalo, the satellite is no longer tracked by the algorithm; full details of the algorithm as well as tests of the approach may be found in \cite{Behroozi2013b}.

\subsection{SUBFIND - D-TREES} 

\subsubsection{Halo finder}

\subfind~\citep{Springel2001}, {similar to} the other halo finders that we use, is a self-{bound} particle substructure finder. \subfind~first identifies parent groups using a standard FoF linking length of $b=0.2$. Then, the main and satellite subhaloes, defined as locally overdense regions  {as explained below}, are extracted from each pre-selected parent group  {as set out below}. 

In order to identify the gravitationally bound subhaloes, a local density is estimated for each particle with adaptive kernel interpolation using a prescribed number of smoothing neighbours,  N$_{\rm dens}$. Therefore, each particle is considered as a tracer of the density field, and any locally overdense region within this field is considered a candidate halo. Then, for each particle, the nearest {N$_{\rm dens}$} neighbours are considered for identifying local overdensities, {defined as a region that is} enclosed by an isodensity contour that traverses a saddle point within the global {density} of the candidate halo. The algorithm uses two free parameters, N$_{\rm dens}$ and N$_{\rm ngb}=20$ which represents the minimum number of particles for identifying a subhalo and  sets the desired mass resolution for halo identification. (N$_{\rm dens}$  typically uses a slightly larger value than N$_{\rm ngb}$.) This procedure is carried out in a top-down fashion, starting from the particle with the highest  density, additional particles {being} added in a sequence of decreasing density. If a particle is only surrounded by denser neighbours in a single candidate halo, it is added to this region. Whenever a saddle point in the global density field is reached such that it connects two disjoint overdense regions, the smaller candidate is treated as a separate satellite subhalo candidate.  

All candidate subhaloes, selected using only spatial information, are then subjected to an iterative unbinding procedure, using a tree-based calculation of the potential. If the number of remaining bound particles is at least N$_{\rm ngb}$, then the candidate is recorded as a subhalo. The initial set of candidate subhaloes forms a nested hierarchy that is processed inside out, allowing the detection of haloes within satellite subhaloes. However, a given particle may only be a member of one {subhalo}, that is, \subfind~decomposes the initial group into a set of disjoint self-bound {subhaloes}.
For all {subhaloes} the particle with the minimum gravitational potential is adopted as the subhalo centre, and the subhalo mass corresponds to the  sum of the masses of the particles associated with them. For the main {subhalo}, \subfind~additionally calculates a SO mass around this centre, but we do not use this mass in our study.

\subsubsection{Tree builder}

The \dtrees~ algorithm attempts to reliably track the most bound cores of halos (and subhalos) despite uncertainty in the definition of the halo boundary and possible loss of particles between simulation snapshots. The algorithm is described in detail in appendix A2 of \citealt{Jiang2014}, so only an overview is included here. Given a pair of simulation snapshots we can identify the most bound core of each halo in one snapshot and determine which halo contains the largest part of it in the other snapshot. This is done by following the 10-100 most bound particles. If we have a progenitor halo A and a descendant halo B such that halo A's most bound core belongs to halo B at the later time and halo B's most bound core came from halo A at the earlier time, then we can be confident that halos A and B are the same object identified at different times and we call halo A the main progenitor of B. Mergers are identified by cases where the progenitor's core goes to the descendant but the descendant's core originated elsewhere.

Merger trees are constructed by applying this method to adjacent pairs of snapshots. In cases where a halo is not the main progenitor of its descendant in the next snapshot we search several subsequent snapshots and attempt to find a descendant for which the halo is the main progenitor. This allows the algorithm to locate descendants in cases where the halo finder temporarily loses track of the halo, such as when a satellite subhalo passes close to the core of its host halo.

\subsection{VELOCIRAPTOR - D-TREES} 

\subsubsection{Halo finder}
\label{velociraptor: halo finder}

\velociraptor~\footnote{\href{https://bitbucket.org/pelahi/velociraptor-stf/}{https://bitbucket.org/pelahi/velociraptor-stf/}} \citep{Elahi2011} is a main and satellite subhalo finder that identifies objects in a two-step process. Haloes are identified using a 3DFoF algorithm and are then fed to 6DFoF \citep{Diemand2006} to prune artificial particle bridges. The 6DFoF algorithm links particles together if they are closer than some distance metric which has an extension to include a proximity condition in velocity space. Two particles are linked if 
    \begin{equation}
    \frac{(x_1 - x_2)^2}{b^2l^2} + \frac{(v_1 - v_2)^2}{\alpha^2\sigma^2} < 1 ,
    \end{equation}
where $bl$ is the real-space linking length, with $b=0.2$ and $l$ the mean interparticle separation in the simulation, and $\alpha \sigma$ is the velocity-space linking length, with $\alpha=1.25$ and $\sigma$ the velocity dispersion of the 3DFoF halo.   
6DFoF is also able to flag major mergers, as two (or more) large phase-space dense cores in the FoF halo. \velociraptor~follows the normal convention and treats the smaller object(s) as a satellite and the larger one as a main subhalo. Further substructures are searched for by identifying particles that appear to be dynamically distinct from the mean halo background, i.e., particles which have a local velocity distribution that differs significantly from the averaged background halo. These dynamically distinct particles are linked with a phase-space FoF algorithm into substructures. This approach is capable of not only finding satellites but also unbound tidal debris surrounding them as well as tidal streams from completely disrupted satellites. For this analysis we only take self-bound groups and use the number of particles in a subhalo to calculate  its mass.  


\subsubsection{Tree builder}

\velociraptor~is accompanied by the tree builder \treefrog~\citep{Elahi2011}. We originally intended to construct merger trees from our \velociraptor\ outputs using \treefrog, but this was not possible due to numerical issues. Instead we used \dtrees\ to build merger trees for \velociraptor\ subhaloes.

\section{GALFORM galaxy formation model}
\label{SAM description}

Here, we use  the  \galform~ SAM introduced by \citet{Cole2000}, with the modifications and improvements to the \citet{Lacey2016} version as presented in \citet{Baugh2019}. 

\galform~ is composed of two main parts, (i) the Dhalo algorithm that processes the {subhalo merger trees} in order to obtain the halo merger trees (the base algorithm is described in \citealt{Helly2003}, and we use the version of \citealt{Jiang2014}), and (ii) the algorithm that takes these trees and follows the baryonic physics associated with them; even though the latter is usually simply referred to as \galform, the model is only complete when the two parts are applied to a simulation.  Therefore, the output of the four merger tree finders presented in the previous section needs to be processed and homogenised first with Dhalo, as we now describe.  

\subsection{Constructing Dhalo Merger Trees}
\label{Constructing Dhalo Merger Trees}

    \begin{figure}
    \includegraphics[scale=0.36]{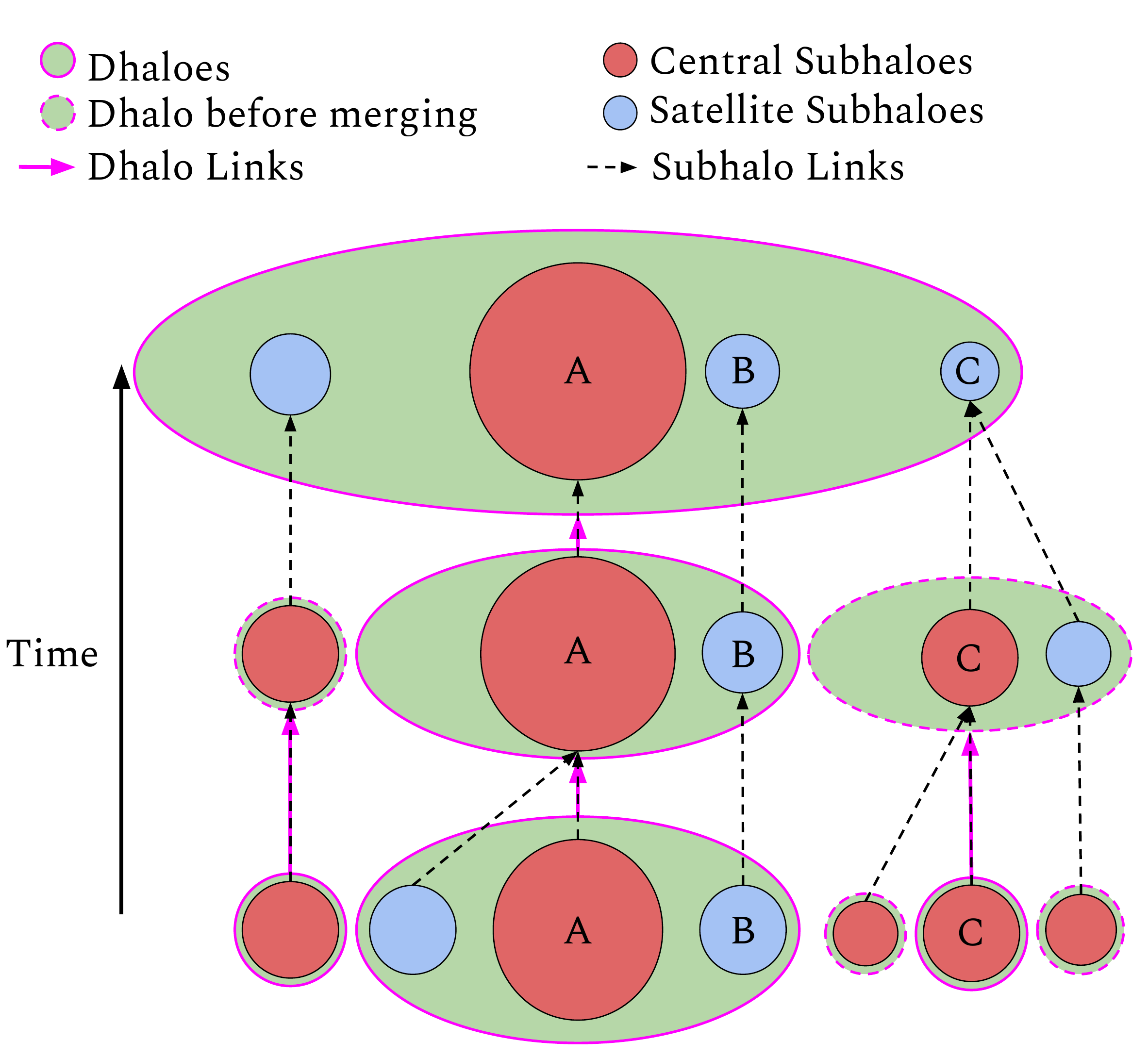}
    \caption{
    Schematic showing Dhalo trees. Red and blue circles denote central and satellite subhaloes, whereas green circles/ellipses show Dhaloes that contain central and satellite subhaloes. The links between subhaloes are shown as black dashed lines, whereas Dhalo links are shown as pink arrows, and they overlap with central subhalo links.  {Dashed circles highlight a Dhalo that is about to merge with another Dhalo.}
    }
    \label{Dhalo_cartoon}
    \end{figure}

The four merger tree builders used here have some similar characteristics. For example, they all use some variant of the FoF algorithm as a starting point. Despite this, the subhalo merger trees generated differ in both the subhalo definition employed and the way in which descendants and progenitors are identified. 
On the other hand, these merger trees are all  used here as inputs to  \galform~. The need for consistency between the halo/subhalo model used in the SAM calculation and the N-body simulation imposes some requirements on the construction of the merger trees.

We use the Dhalo algorithm described in  \citet{Jiang2014} to post-process the subhalo merger trees generated by the tree builders \hbt, \rockstart, \subfindt~and \velociraptort. The {Dhalo} algorithm groups subhalo merger tree branches together to form dark matter haloes whose  {masses increase monotonically} and which avoid temporary mergers due to tenuous bridges of particles or an overlap of their diffuse outer haloes. These haloes are thus well suited for modelling galaxy formation, and their merger trees form the basis of \galform. The {Dhalo} code is also used to convert the merger trees into the format required by \galform.

Below we  give an outline of the Dhalo algorithm; a full description can be found in Appendix A3 of \citet{Jiang2014}.

\subsubsection{Halo Catalogue Input}

The first step in building merger trees is the construction of catalogues of main subhaloes and their satellite subhaloes, as identified by \hbt, \rockstar, \subfind~and \velociraptor. Dhalo does not apply modifications to these subhalo catalogues, i.e. it uses subhalo masses and subhalo merger trees as provided by the subhalo tree finder. However, it can change main subhaloes to satellites and vice versa compared to the original definition as we will see  below.   

\subsubsection{Building the {Dhalo} trees}

After identifying the main and satellite subhaloes from each halo finder, the Dhalo algorithm processes these subhalo merger trees as follows. It partitions the merger trees into discrete branches. A new branch begins whenever a new Dhalo forms and continues for as long as the Dhalo exists in the simulation. When a merger occurs, the Dhalo algorithm decides which of the progenitor Dhaloes survives the merger by determining which progenitor contributed the largest mass in bound particles of the descendant. The Dhalo branch corresponding to this progenitor continues, while the other progenitor’s Dhalo branch ends.

Halo mergers are typically not instantaneous. An infalling subhalo may pass through the host halo and go beyond the virial radius before falling in again. The Dhalo algorithm deems such objects to have merged with the host halo once they have first lost at least 25 per cent of their mass and are within twice the half mass radius of the host halo. At all later times, the infalling subhalo is considered to be a satellite subhalo, even if it is outside the virial radius. When a Dhalo includes satellite subhalos at large radii this indicates that these satellites passed through the central halo at an earlier time.

Finally, the algorithm defines collections of subhaloes embedded hierarchically within each other as a single `Dhalo', but excludes neighbouring subhaloes that may be part of the same FoF group, but which are only linked by a bridge of low-density material or subhaloes that are beginning the process of merging but have not yet lost a significant amount of mass. Subhaloes are grouped into `Dhaloes' in such a way that once a subhalo becomes part of a Dhalo, it remains a component of that Dhalo’s descendants at all later times at which the halo survives, even if it is a satellite component  temporarily outside the corresponding FoF halo (i.e. it could be classified as a main subhalo by the original halo finder).  All of a Dhalo’s subhaloes which survive at a later snapshot must (by construction) belong to the same Dhalo at that snapshot. We take this to be the descendant of the Dhalo. This defines the Dhalo merger trees.  Fig.~\ref{Dhalo_cartoon} shows a schematic  representation of Dhaloes at different time steps with time increasing towards the top of the figure. Note that Dhalo links are only present when the descendant of the Dhalo is on the same Dhalo branch, rather than having merged with another Dhalo and become a satellite. Some subhaloes are labelled to help interpret the figure. Subhalo A, for instance, is the central subhalo of its host Dhalo in all snapshots.  Subhalo B is always a satellite of the same Dhalo. Subhalo C starts as a central subhalo of a Dhalo with no satellite subhaloes, then acquires a satellite, and finally becomes a satellite subhalo of a Dhalo in the final snapshot. 

Central and satellite subhaloes of a Dhalo are defined as follows. By default the central subhalo is the one with the most mass in its past merger tree {starting from the latest snapshot at which the Dhalo existed in the simulation}. This avoids the centre switching between different subhaloes as they fluctuate in mass over time. All other subhaloes are treated as satellites.   {All satellite subhaloes that are resolved at any given snapshot are referred to as type 1 satellites.}

Initially the Dhalo mass is set equal to the sum of the masses of the subhaloes assigned to the Dhalo. \galform~ then forces each Dhalo to have at least as much mass as the sum of its progenitors at the previous output time by adding mass to the current Dhalo where necessary. This is done starting at early times and working forwards, because adding mass at one time can cause a later Dhalo to be less massive than its progenitors. 

The Dhalo masses that have been forced to increase monotonically in time in this way are used by \galform~to calculate the evolution of the baryonic components of galaxies.

\subsubsection{Type 1 and 2 satellite subhaloes}

Satellite subhalos that are still identified in the simulation are referred to as type 1. None of the finders except \hbt~ keep track of the satellite subhaloes that have completely merged into other subhaloes. For those finders, when a satellite subhalo can no longer be detected in the simulation, the subhalo merger tree will show that it merged with the main subhalo.  {This type of subhalo is referred to as a type 2 satellite.} 

In \galform~ it is assumed that this merger is due to the limited resolution of the simulation and that the subhalo should still exist for some time after that. Because of this, when we describe the statistics of subhaloes and merger trees in Section \ref{section: Merger Trees} we will also show properties of the subhaloes that have already merged.  In \hbt~ no extra work is needed for this; in the other finders it is a matter of traversing back in time along the merger tree to find all merged satellites. In particular, we will look at their abundances, their Dhalo mass before they became satellites, and the ratio of their subhalo mass to that of the Dhalo within which they are merging. The latter quantity is used by \galform~ to estimate the extra time the galaxy will take {after the disappearance of the subhalo} to finally merge with the central galaxy of the Dhalo. During that time any galaxy associated with the subhalo is placed on what was the most bound particle of the subhalo when it last existed.

From this point forward  we will refer to the steps described in this section as the postprocessing of merger trees by \galform.

\subsection{Matched subhaloes}
\label{ssec:match}

Throughout we will need a matching procedure between subhaloes (either satellite or main) in the catalogues resulting from the different finders. To do this we search for subhalos that satisfy the following two criteria:
    \begin{itemize}
    \item Subhalo positions within $30$ per cent of the half mass radius of subhaloes of a different finder.
    \item Subhalo masses from the different finders within a factor 3 of each other.
    \end{itemize}
If there is more than one match we pick the least distant one. When applying this  procedure to \subfind~ and \hbt, it allows us to match almost $100\%$ of the subhaloes (main or satellite) of \subfind~with \hbt~subhaloes for subhalo masses above $10^{10}~h^{-1} M_{\odot}$. In less than 1 in $1000$ cases we find more than one {possible} match for any given subhalo {before choosing the least distant one}, and in a percentage that increases for lower subhalo masses, {main} subhaloes of one finder are matched to satellite subhaloes of the other.  

When applying this matching procedure to catalogues from \hbt, \rockstar~and \subfind~we find matches for almost $100$ per cent of the subhaloes. \velociraptor, however,  yields more satellite subhaloes than the other finders and this leads to a lower rate of matches, dropping to about $90$ per cent for subhalo masses above $10^{10} h^{-1} M_{\odot}$. The matching procedure consistently returns more than one match for about 1 in $1000$ subhaloes {before choosing the least distant one}. This rate increases to about 1 in a $100$ when the match is done using only satellite subhaloes.

\subsection{Baryonic physics in GALFORM}

Here we briefly summarize the baryonic physics implemented in \galform. Further details can be found in \citet{Lacey2016} and \citet{Baugh2019}.

SAMs use simple, physically motivated equations to follow the fate of baryons in a universe in which structure grows hierarchically through gravitational instability (for an overview see \citealt{Baugh2006} and \citealt{Benson2010b}).  \galform~ models the main physical processes that shape the formation and evolution of galaxies, such as 1) the collapse and merging of dark matter haloes, 2) the shock heating and radiative cooling of gas inside dark matter haloes, leading to the formation of galactic discs, 3) quiescent star formation (SF) in galaxy discs, 4) feedback from supernovae (SNe), from heating by active galactic nuclei (AGN) and from photoionization of the intergalactic medium, 5) chemical enrichment of stars and gas and 6) galaxy mergers driven by dynamical friction within common dark matter haloes which can trigger bursts of SF, and lead to the formation of spheroids.

The model includes growth of supermassive black holes by accretion of gas during starbursts and directly from the hot halo, and by mergers of black holes \citep{Bower2006,Fanidakis2012,Griffin2019}, and  the improved treatment of SF implemented by \citet{Lagos2011}. This latter extension splits the hydrogen content of the ISM into atomic and molecular hydrogen. The model allows bulges to grow through minor and major galaxy mergers and through global disc instabilities.  An improvement over previous versions of the code is that \galform~now follows the resolved satellite subhalos down  to the moment when they are lost within the central subhalo and calculates a dynamical friction timescale from the last time at which the satellite subhalo was identified, and assumes that the galaxy merges into the central galaxy after this timescale \citep{Simha&Cole2018}. Previous versions of the code merged the satellite galaxy with the central galaxy as soon as the dynamical friction timescale, calculated at the time the galaxy became a satellite, was exhausted,  { even if the corresponding subhalo  can still be resolved in the simulation.  In analogy with the terminology used for satellite subhaloes, satellite galaxies hosted by a {resolved} subhalo (which could have been a central subhalo in earlier simulation time-steps) will be referred to as type 1 satellite galaxies. Conversely, a type 2 galaxy satellite, is a satellite galaxy that is not associated to a resolved satellite subhalo at the present timestep, but at an earlier timestep was hosted by a satellite or central subhalo; at the present timestep the latter can no longer be identified by the halo finder due to proximity to the central subhalo center or due to its disruption. }  

    \begin{figure*}
    \includegraphics[scale=0.4]{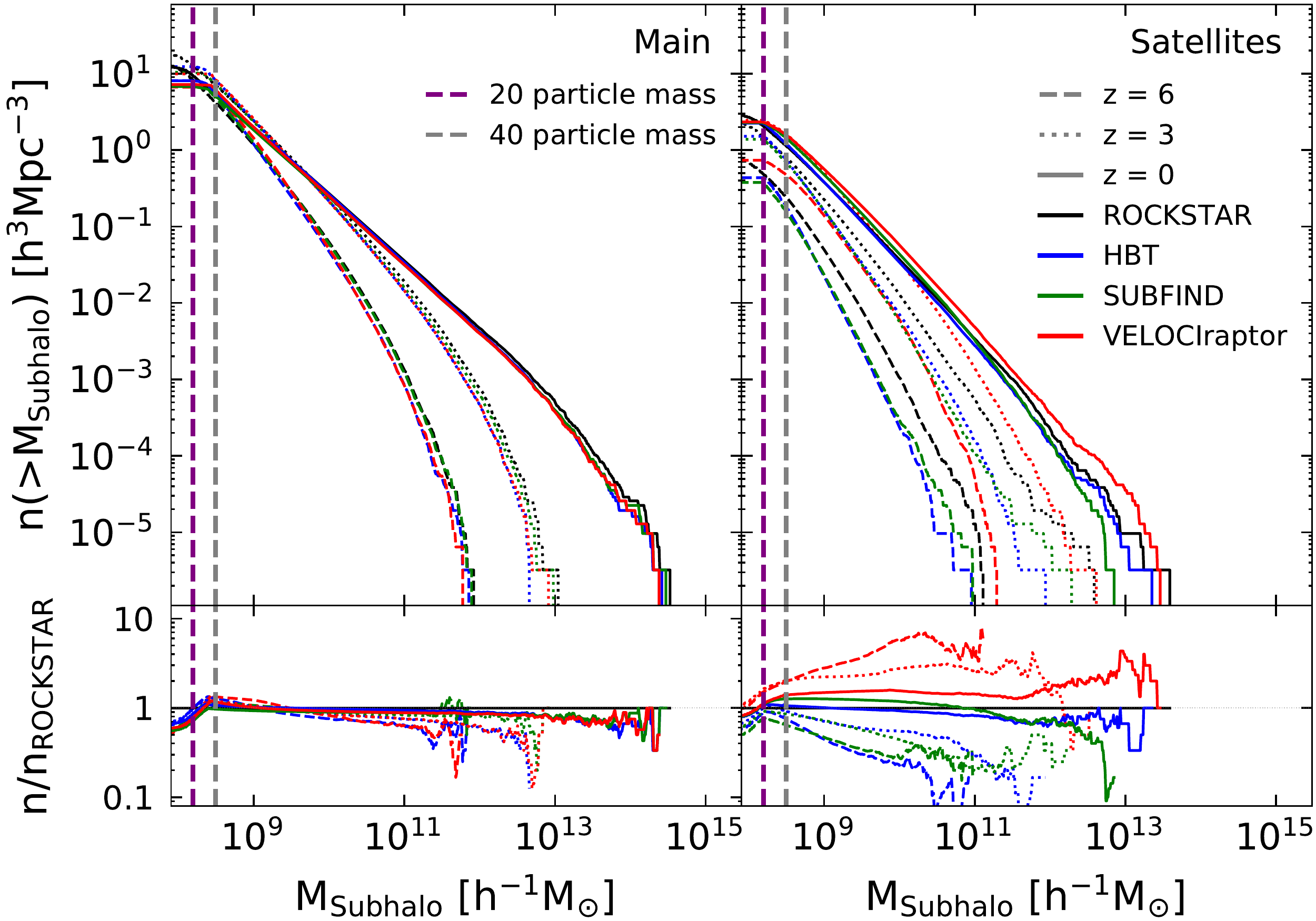}
    \caption{Cumulative mass function of main subhaloes (left) and satellite subhaloes (right), at redshift $z=0$, $z=3$ and $z=6$ for the four halo finders, as labelled. The subhalo masses are those prior to being processed by Dhalos or \galform. The vertical purple dashed line indicates the mass corresponding to 20 particles; the vertical gray dashed line indicates the  mass corresponding to 40 particles where the cumulative mass function starts to depart from a power law; this is taken as the completeness limit. The lower panels show the ratios of the cumulative mass functions of the three other finders with respect to \rockstar.}
    \label{fig: cshmf}
    \end{figure*}
    
    \begin{figure*}
    \includegraphics[scale=0.55]{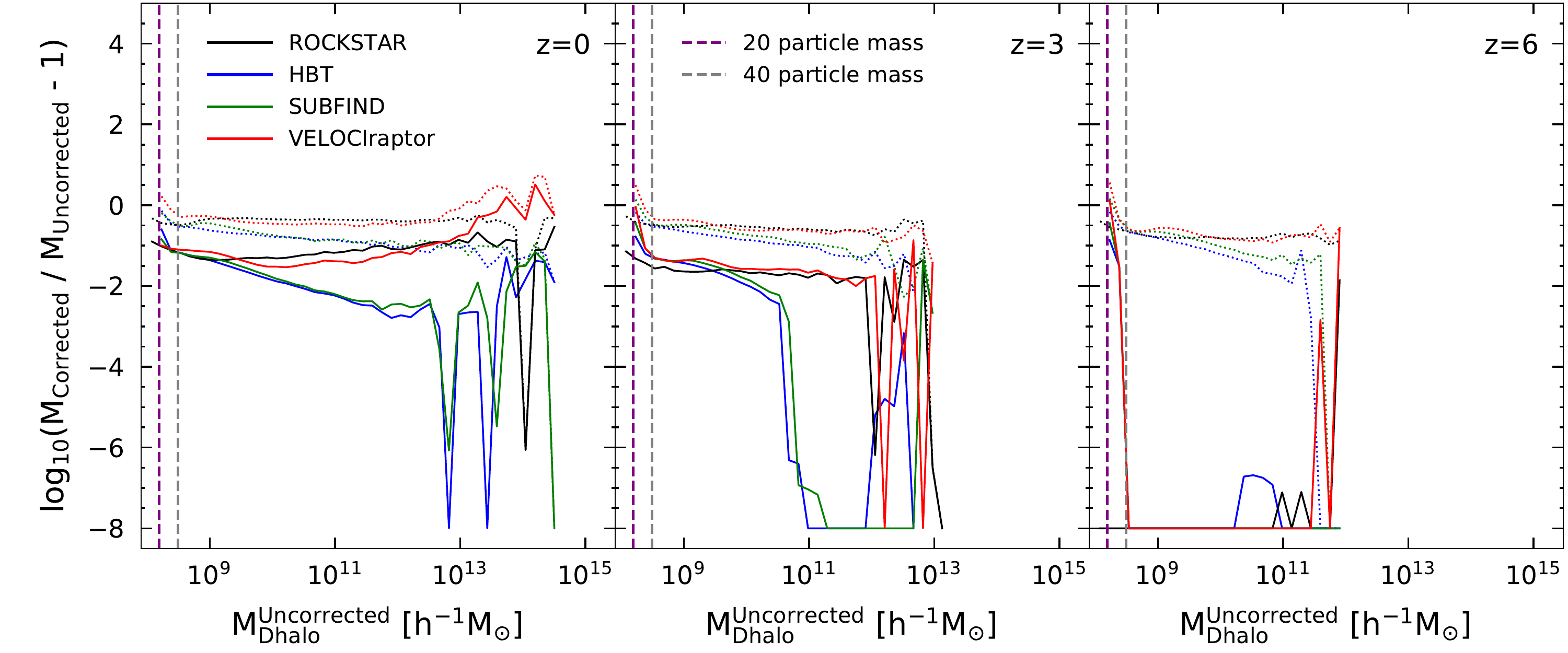}
    \caption{Median fractional change (solid lines) in Dhalo mass resulting from the monotonicity requirement of GALFORM as a function of the uncorrected Dhalo mass at redshift $z=0$, $z=3$ and $z=6$ (left, middle and right). The 90 percentile is shown as dotted lines. We artificially set the excess to $10^{-8}$ when the {constraint} produces no change in the Dhalo mass.}
    \label{fig: dhalo ratio = corrected/uncorrected}
    \end{figure*}
    
    \begin{figure*}
    \includegraphics[scale=0.4]{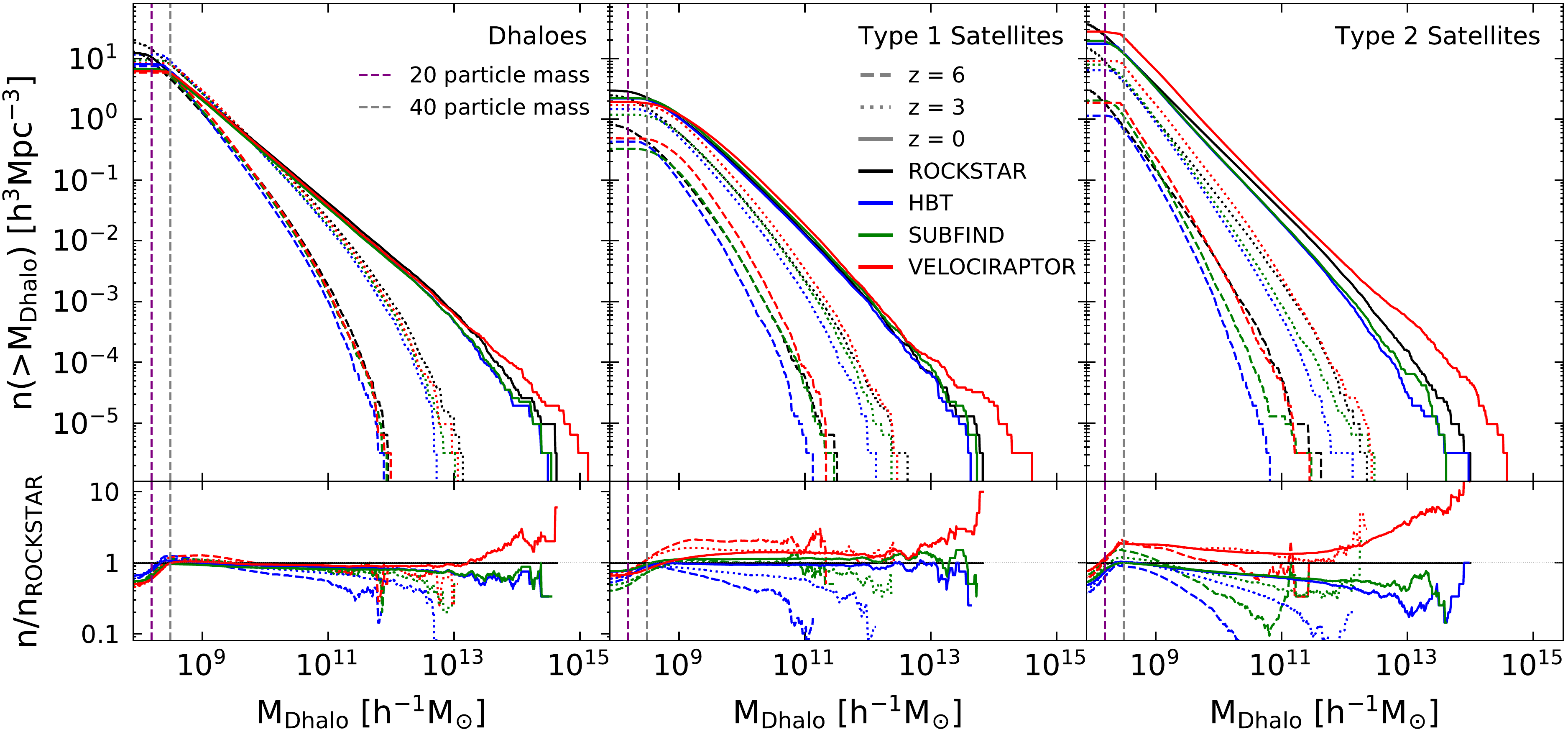}
    \caption{
    Cumulative mass function of corrected Dhaloes (left), and infall Dhalo masses of type 1 and type 2 satellite subhaloes (centre and right panels, respectively). The cumulative mass functions are shown for redshifts $z=0$, $z=3$ and $z=6$ for the four halo finders, as labelled. These Dhalo and infall masses correspond to masses after being processed by Dhalos {and} \galform. The centre and right panels show the infall mass for the satellites at the last snapshot before they became satellites. The vertical purple and gray dashed line are the same as in Fig. \ref{fig: cshmf} {and show mass limits corresponding to 20 and 40 particles}. The lower panels show the ratios of the cumulative mass functions of the three other finders with respect to \rockstar.
    }
    \label{fig: cdhmf}
    \end{figure*}

    \begin{figure*}
    \centering
    \includegraphics[scale=0.173]{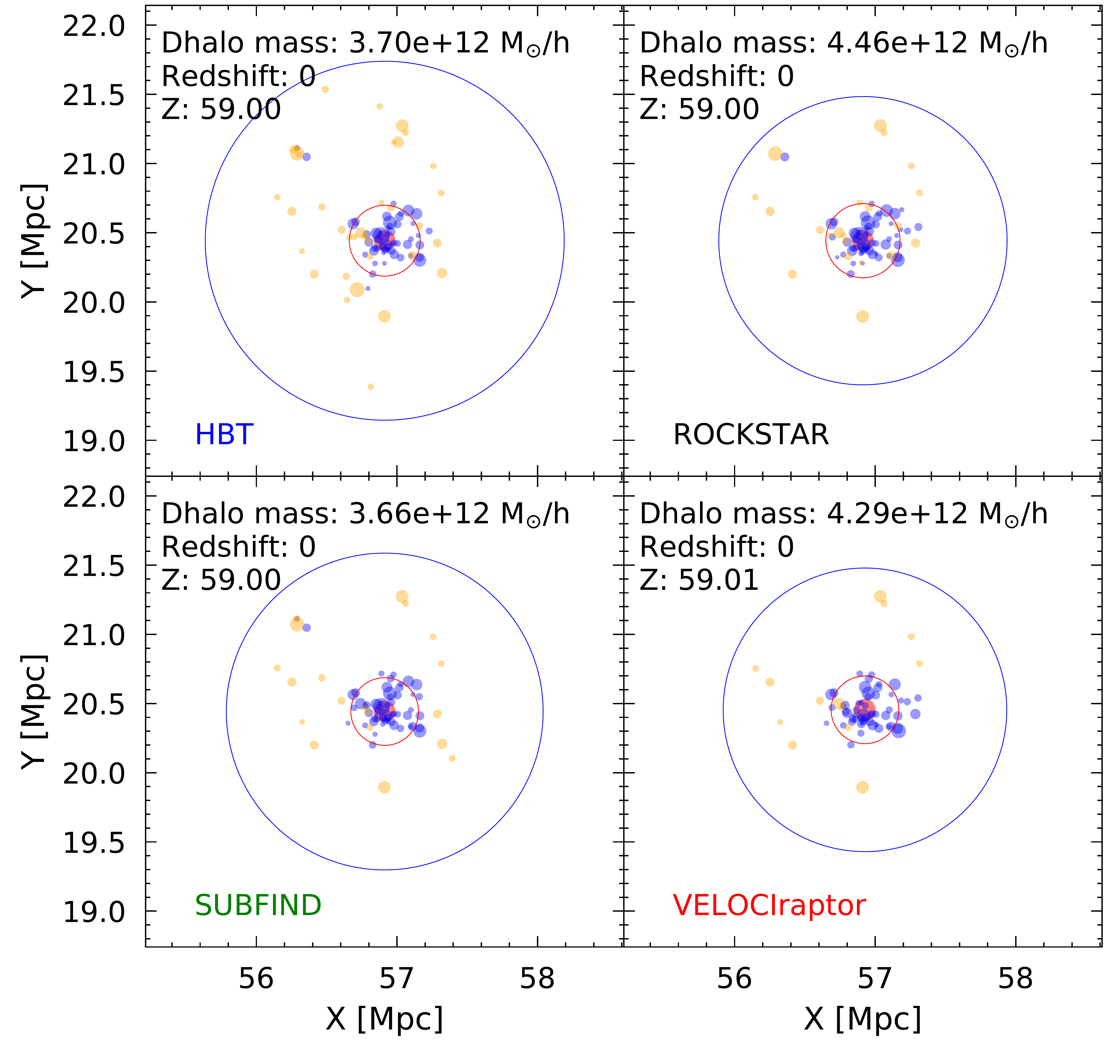}
    \includegraphics[scale=0.173]{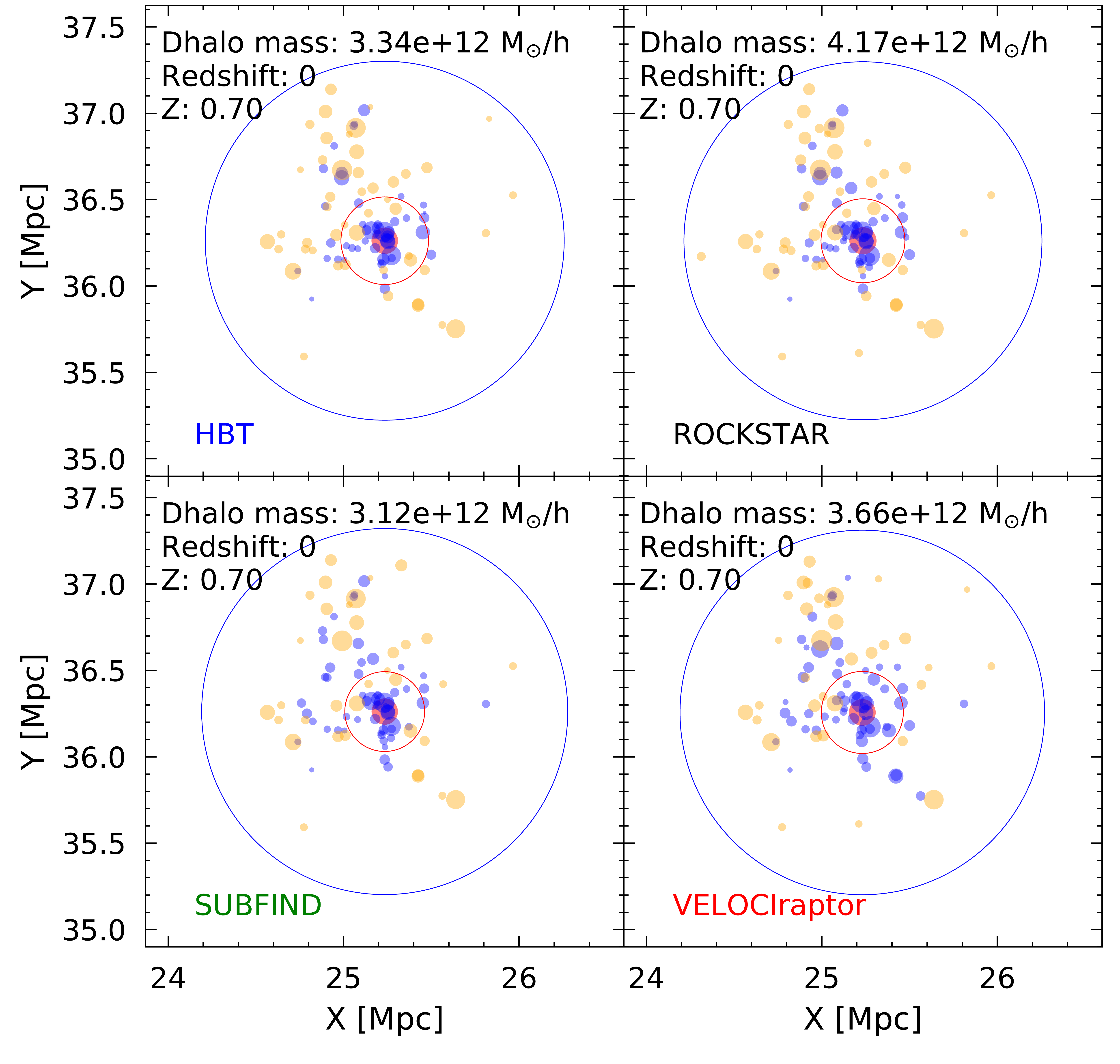}
    \includegraphics[scale=0.173]{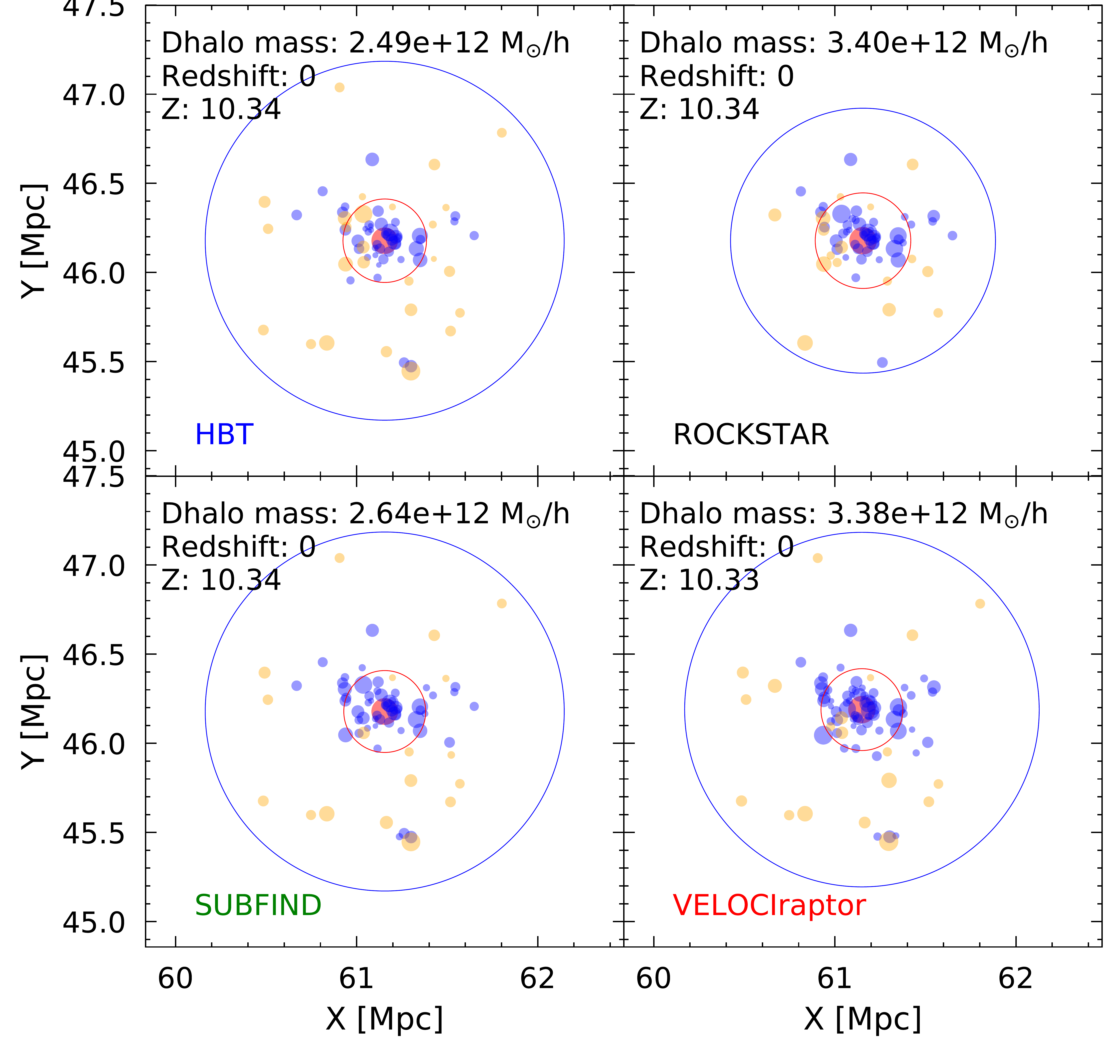}
    \includegraphics[scale=0.173]{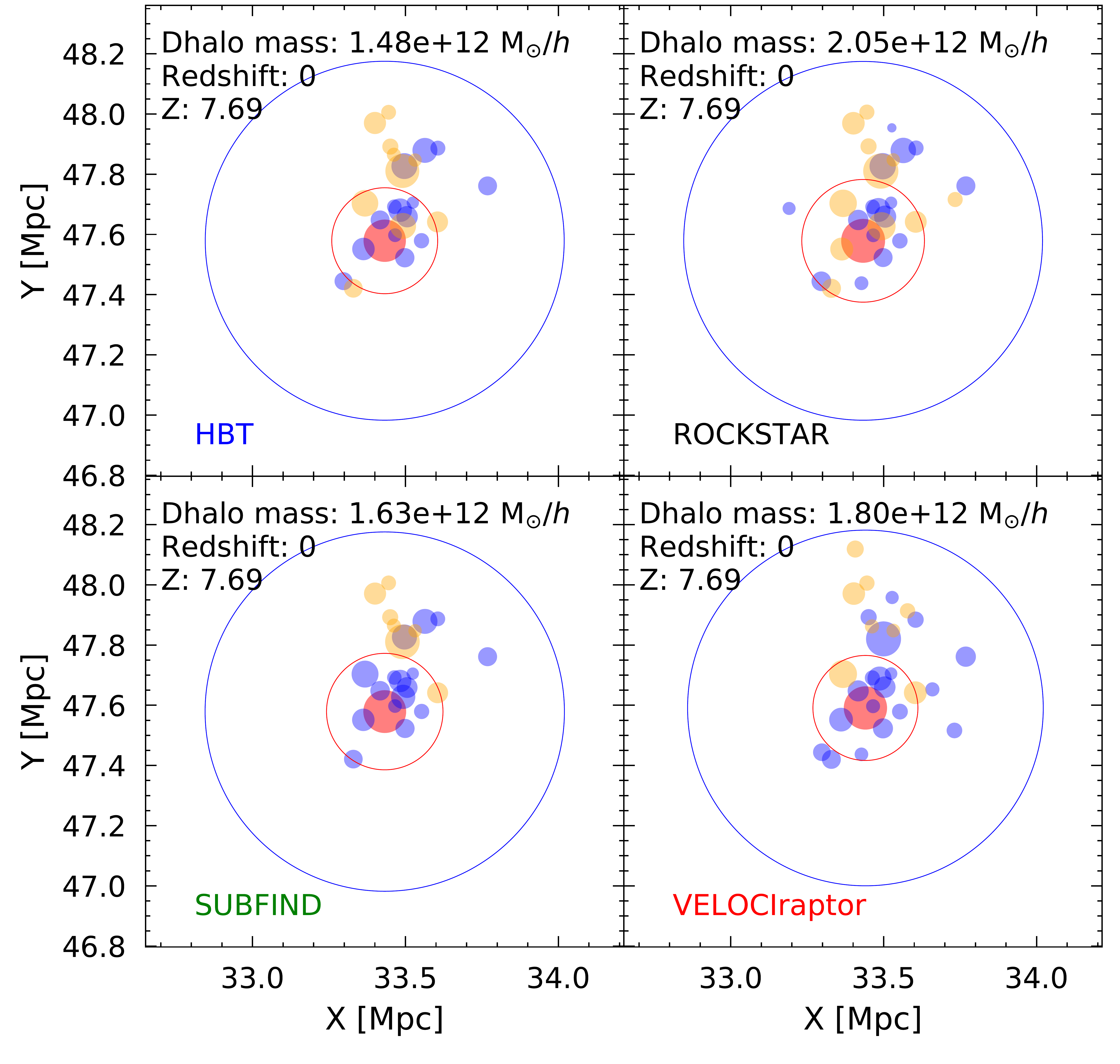}
    \caption{XY-projections of matched Dhaloes of similar mass (solid, light red circles) found  in all four halo finders by matching their positions { and masses as explained in Sec. \ref{ssec:match}. Each set of four panels correspond to four different Dhaloes; subpanels in each set show the matched Dhalo as found by the different halo finders.}  Solid blue circles show  type 1 satellite subhaloes and solid yellow circles show other Dhaloes, with Dhalo mass greater than $3.12\times 10^8$\Mh, within {the} sphere bounded by the largest blue circle. The size of the circles is proportional to  $\log$(M/\Mh) of the Dhalo mass, or infall mass for type 1 satellite subhaloes. The solid blue circles enclose the farthest satellite and the filled red circles correspond to twice the half mass radius of the central Dhalo.}
    \label{fig: postion examples}
    \end{figure*}

    \begin{figure*}
    \centering
    \includegraphics[scale=0.55]{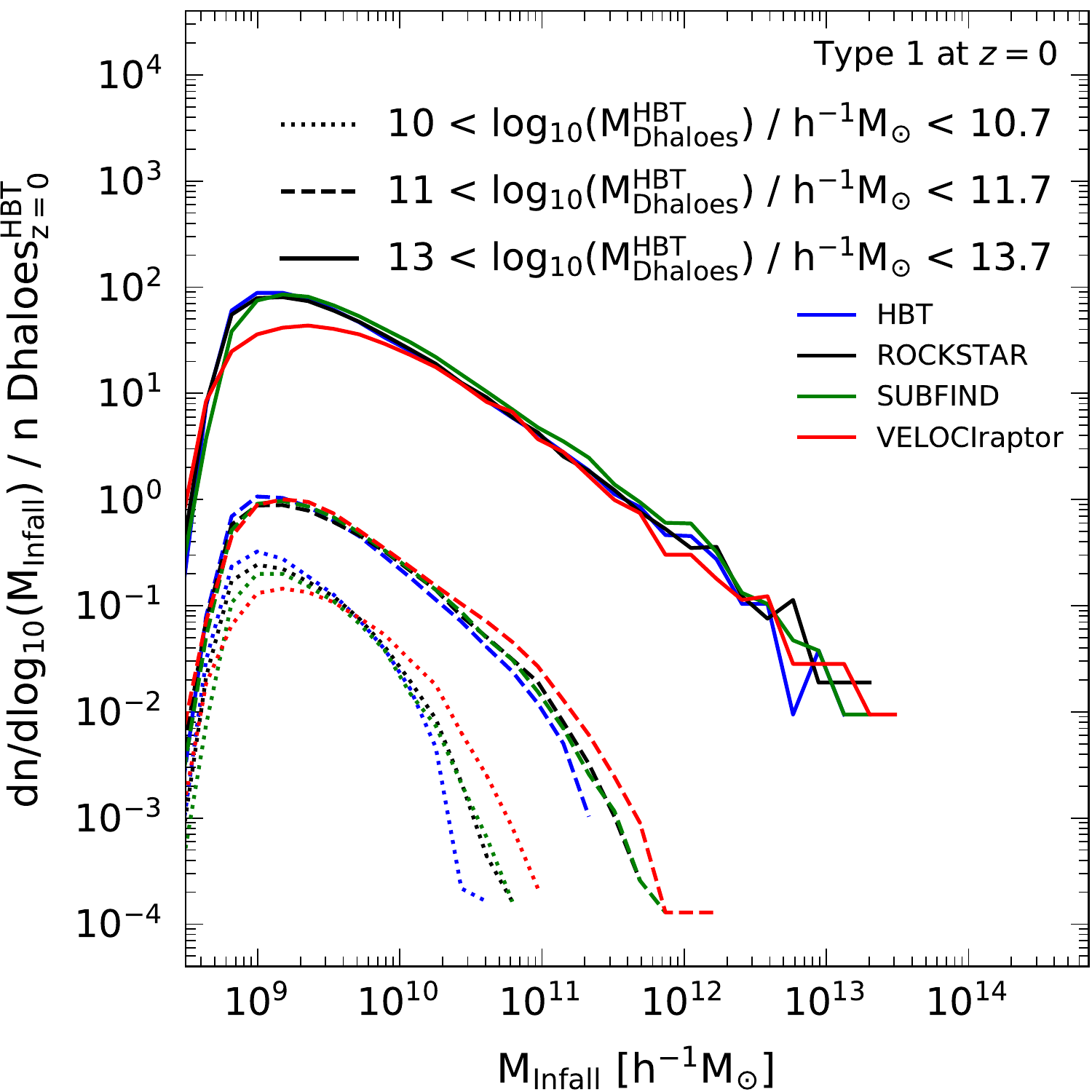} \hspace{0.2cm}
    \includegraphics[scale=0.55]{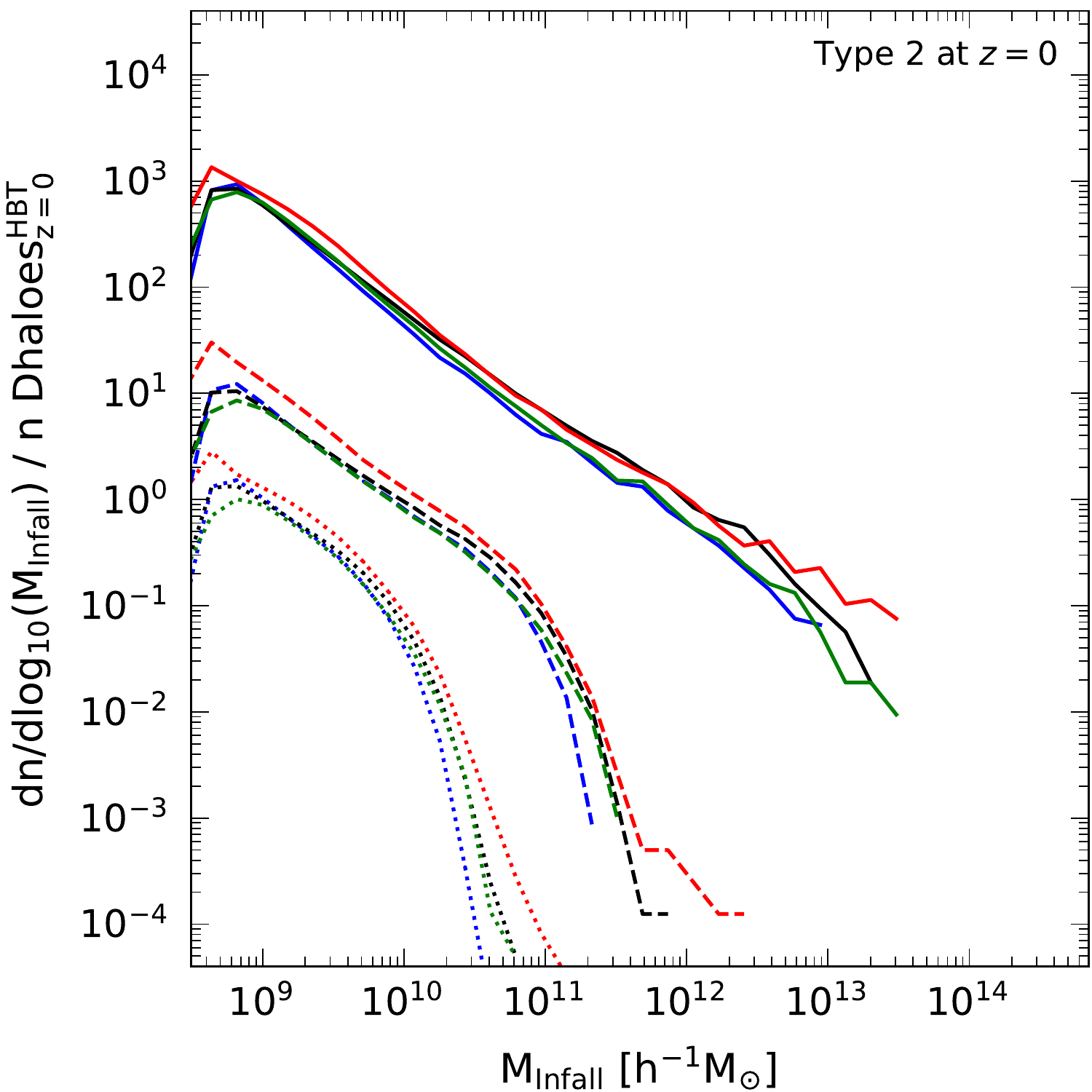}
    \caption{{Mass function of type~1 and type~2 satellite subhaloes at redshift $z=0$ for the four halo finders, as labelled}. Left panel shows all surviving satellite subhaloes (type~1 satellites), whereas the right panel shows this only for type~2 subhalos, i.e. satellite subhaloes that have merged with their main subhaloes. {The mass functions are based on Dhalo infall mass for the satellites at the last snapshot before they became satellites. Type~1 and type~2 satellite subhaloes shown were selected from} matched $z=0$ Dhaloes found in all four halo finders. The mass functions are shown for 3 different Dhalo infall mass ranges, represented by different line styles, using the \hbt~mass as a reference. The y-axis is normalized by the number of matched \hbt~Dhaloes at $z=0$ found in each Dhalo infall mass range.}
    \label{fig: type 1 and 2 satellites subhaloes}
    \end{figure*}

    \begin{figure}
    \centering
    \includegraphics[scale=0.55]{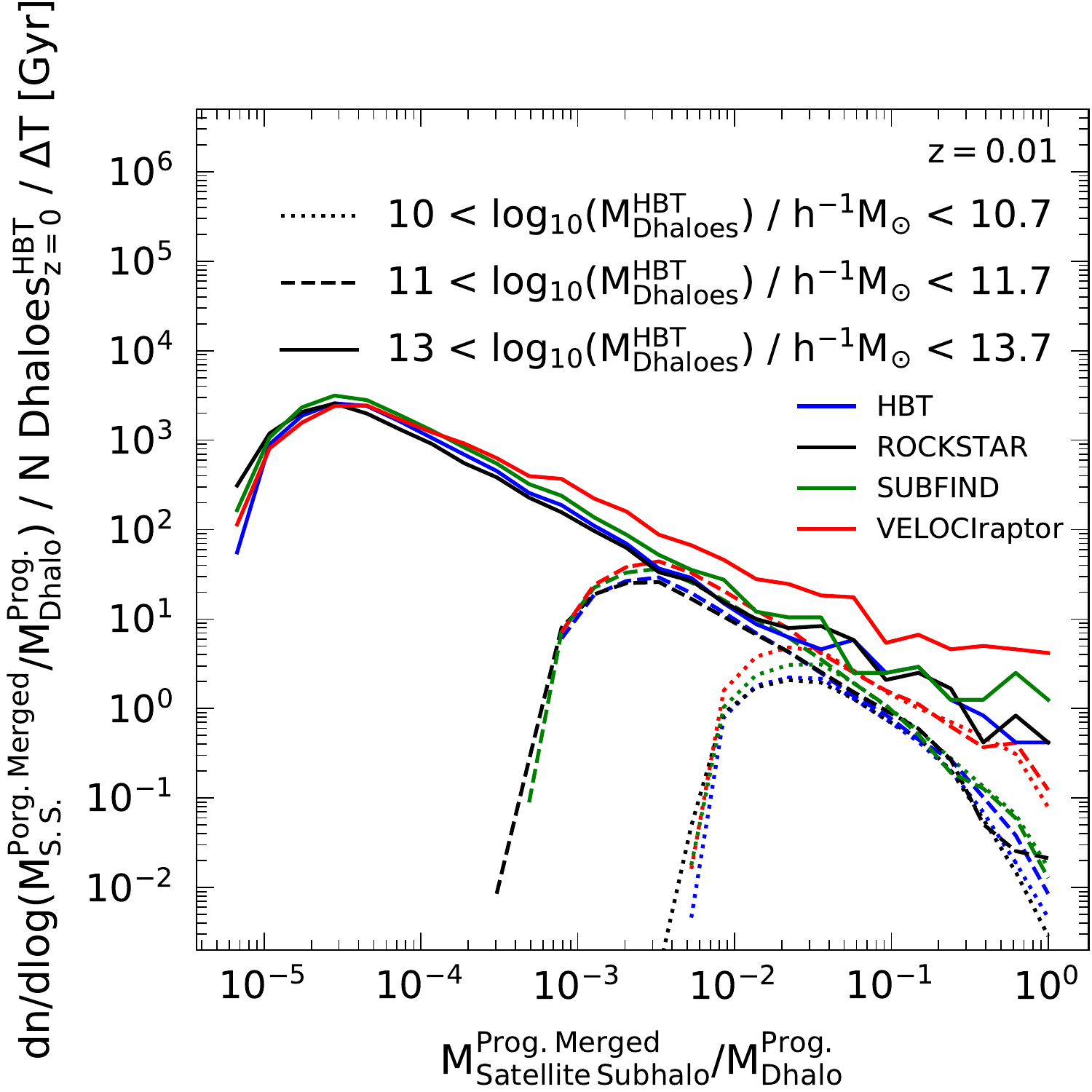}
    \caption{
    Distribution of the mass ratio between satellite subhaloes and the host Dhaloes within which they merged with the central subhalo, for subhalo mergers occuring between $z=0.01$ and $z=0$. All mass ratio quantities are calculated from the $z=0.01$ Dhalo progenitor of the matched $z=0$ Dhalo samples for 3 ranges of $z=0$ \hbt~Dhalo masses as reference. The y-axis is normalized by the number of Dhaloes in each mass range at $z=0$
    and the time interval $\Delta T = T(z=0)-T(z=0.01)$ over which we sample subhalo mergers.
    }
    \label{fig: ratio sat1/dhalo}
    \end{figure}

\section{Comparison of Halo Merger Trees}
\label{section: Merger Trees}

Halo merger trees are the backbone of SAMs. In \galform, galaxy properties are calculated from prescriptions directly related to the properties of the dark matter haloes and their evolution. Therefore, in this section we present a comparison of merger trees and the resulting evolution of dark matter haloes between the different merger tree builders described in Section  \ref{merger_tree_builder_descrip}. 

\subsection{Differences in halo mass functions}

The halo finders described in Section \ref{merger_tree_builder_descrip} assign particles to subhaloes in different ways and this results in different masses, even after their outputs are postprocessed and converted into the Dhalo format (Section \ref{Constructing Dhalo Merger Trees}).  

Fig.~\ref{fig: cshmf} shows the {cumulative mass functions of main and satellite subhaloes for the different halo finders (colours, indicated in the key) at three different redshifts, $z$=0, 3 and 6 (left, centre and right, respectively), before any processing by Dhalos or \galform}.  {In all cases apart from \rockstar\ the masses correspond to the number of particles in each object, multiplied by the particle mass.} This allows us to {directly} compare the different finders at this stage. {The cumulative mass functions resulting from different finders are very similar for main subhaloes (Fig.~\ref{fig: cshmf}, left), with differences of only about $20-30$ percent.  The differences are larger for the satellite subhalo mass functions (Fig.~\ref{fig: cshmf}, right) but they are still  similar qualitatively.  In general, the amplitude of the cumulative mass function for subhaloes decreases with increasing redshift}. 

For main subhalo masses (Fig.~\ref{fig: cshmf},  left), {we see that} differences between halo finders increase with increasing redshift. {The mass functions of main subhalos generally show} larger abundances at a fixed halo mass for \rockstar, and similar lower abundances for the other three finders, reaching the largest difference for the highest halo masses. {However, these differences are modest, less than a factor $\sim 2$ over the ranges probed here}. {The right panel shows mass functions of satellite subhaloes, for which the differences between halo finders are seen to be larger than for main subhaloes}, with the highest abundances for \velociraptor~ which {shows} satellite abundances higher by up to a factor of $\sim 7$ ($\sim 2$) compared to \rockstar~ at $z=6$ ($z=0$). The lowest {satellite} abundances are returned by \hbt.

The cumulative mass functions are seen to flatten at low masses, reflecting the resolution of the simulation and the minimum particle number set in the finders. The lower mass limits for detected subhaloes {differ between halo finders}. \subfind~and \velociraptor~detect subhaloes above $M_{\rm halo}=1.56\times10^8$\Mh, corresponding to 20 particles, the lower limit set in those halo finders (purple vertical line). Although \hbt~is also built to find main and satellite subhaloes {with} at least 20 particles, it also maintains a record of type 2 satellite subhaloes as having a single dark matter particle to indicate they {have} already merged (not visible in the figures); we {gave} more details about this feature of \hbt~in Section \ref{subsubsection: HBTtree}. For \rockstar~the lower mass limit {is} $M_{\rm subhalo}=1.56\times10^7$\Mh, {corresponding to} 2 dark matter particles (see \S~\ref{subsubsection: rockstar}).

It can be seen that {different halo finders vary in their} ability to resolve subhaloes in the mass range $M_{\rm subhalo} \sim 1.56$ -- $3.12\times 10^8$\Mh, i.e., the mass corresponding to $20$ to $40$ particles (the latter is marked as a vertical grey line in Figs. 2, 3 \& 4). The 3D halo finders, \subfind~and \hbt, are able to {find} subhaloes containing as few as 20 particles, but their cumulative mass functions for main subhaloes flatten below 40 particles, which is a sign of incompleteness, particularly at $z=0$. Unlike the 3D halo finders, \rockstar~shows no flattening in the mass function before reaching a mass corresponding to 20 particles. \velociraptor, the other 6D halo finder, also maintains an increasing trend in the cumulative mass function below $40$ particles mass, but not all the way down to $20$ as is the case for \rockstar. The ability of 6D halo finders to resolve lower mass subhaloes has also been reported in \citet{Knebe2011, Knebe2013, Behroozi2015}. This is also the case for low-mass subhaloes in overdense regions \citep{Elahi2011, Onions2013}. Because of this, in order to have a similar completeness among the different halo finders, we impose a lower limit {on} subhalo mass equivalent to $40$ dark matter particles {before processing through  Dhalos and \galform}.

We quantify the effect of the postprocessing of {Dhalo masses} by \galform~ in Fig. \ref{fig: dhalo ratio = corrected/uncorrected}, where we show the change in mass introduced by {the Dhalo mass correction} procedure {forcing the Dhalo masses to increase monotonically in time}, for the different finders and redshifts. {The difference decreases with increasing redshift for all finders}. The median change in Dhalo mass is always below $\sim 10$ per cent for \subfind~and \hbt, and it is smaller with increasing halo mass, at all redshifts. For \rockstar~it is around $10$ per cent and constant with mass at $z=0$, and smaller for higher redshifts, {while for \velociraptor~the effect can reach a factor 3 at high masses at $z$=3}. It is worth noting that the mass increase is similar for \subfind~ and \hbt. The increase is also similar for \rockstar~and \velociraptor, except for the much larger increase for \velociraptor~ at the very highest masses at $z=0$.

This mass increase has some influence on the resulting {cumulative} Dhalo mass functions shown in Fig. \ref{fig: cdhmf}. The left panel shows the abundances of Dhaloes, which can be compared to that of the main subhaloes of Fig. \ref{fig: cshmf}, since in practice most of the mass of the Dhalo is usually in the central subhalo. It can be seen that the differences between {halo finders for} the mass functions of Dhaloes are similar to those found for main subhaloes of \rockstar, \subfind~ and \hbt, but in the case of \velociraptor~ much larger differences are seen {for Dhalo mass functions}, consistent with the mass {increases seen in} Fig. \ref{fig: dhalo ratio = corrected/uncorrected}. The centre and right panels {of Fig. \ref{fig: cdhmf}} show the cumulative mass function of the Dhalo mass of satellites at the last time they were still a main subhalo, i.e. the Dhalo mass at infall. The centre panel shows surviving satellite subhaloes (referred to as type 1 satellites), while the right panel shows satellites that have already merged with the main subhalo (type 2 satellites).
We choose to show Dhalo mass at infall as this quantity is used by \galform\ to calculate satellite galaxy properties. Satellite subhalo masses {are} affected by physical processes such as tidal stripping, but also {by} numerical effects due to the high density environment within Dhaloes, whereas the Dhalo mass at infall is free from these effects. 

{The differences between the number of Dhaloes for different finders in Fig.~\ref{fig: cdhmf} show a similar amplitude as the differences in main subhalo mass functions in Fig.~\ref{fig: cshmf}.} At $z=0$, the algorithms find similar number of Dhaloes and main subhaloes at low masses, {both} with differences within $30$ per cent {between different finders} in the range of low masses, $M_{\rm halo}\sim 3\times 10^8$-$10^{10}$\Mh. 
On the other hand, the abundances of satellite subhaloes as a function of their Dhalo mass at infall show larger differences among the four algorithms {compared to the mass function of Dhaloes}, especially for type~2 satellites. 
For  type 1 satellite subhaloes in Fig.~\ref{fig: cdhmf}  {the differences are smaller than for satellite subhaloes prior to the postprocessing performed by \dhalo~ (cf. Fig.~\ref{fig: cshmf}), particularly at high redshift. This shows that the postprocessing of trees by Dhaloes   reduces the differences between finders for {type 1} satellite subhaloes from a factor of a few seen for satellite subhaloes  to less than a factor of $2$}. 
\subfind\ and \rockstar\ show very similar type~1 satellite subhalo mass functions. 
At $z=0$ \velociraptor\ type~1 satellites are more abundant than those found by \rockstar, \subfind and \hbt at all Dhalo masses. {At high redshifts,} the least abundant type~1 satellites are those of \hbt. 
{In the case of type 2 satellite subhaloes,} \velociraptor\ shows higher abundances than the other halo finders, with the smallest difference at $z=0$, where type 2 satellite subhaloes are up to a factor of $\sim 10$ more abundant than for \rockstar, and even higher with respect to the other two finders. \subfind~type 2 satellites at $z=0$ have similar abundances to those from \hbt. {As was the case with type 1 satellites, at high redshifts, the least abundant type 2 satellites are those of \hbt.}

\subsection{Differences in the definition of main and satellite subhaloes}

The definition of main and satellite subhaloes in each halo finder algorithm plays a crucial role in their identification in the dark matter only simulation. While some algorithms may be able to find subhaloes that another finder misses, they {may} also find the same subhaloes but assign them a different hierarchical classification, e.g., a main subhalo according to one halo finder could be labelled as a satellite subhalo by another one. {For example, } \subfind~finds  more satellite subhaloes than \hbt, which classifies at least some of these as main subhaloes.

    \begin{figure*}
    \includegraphics[scale=0.6]{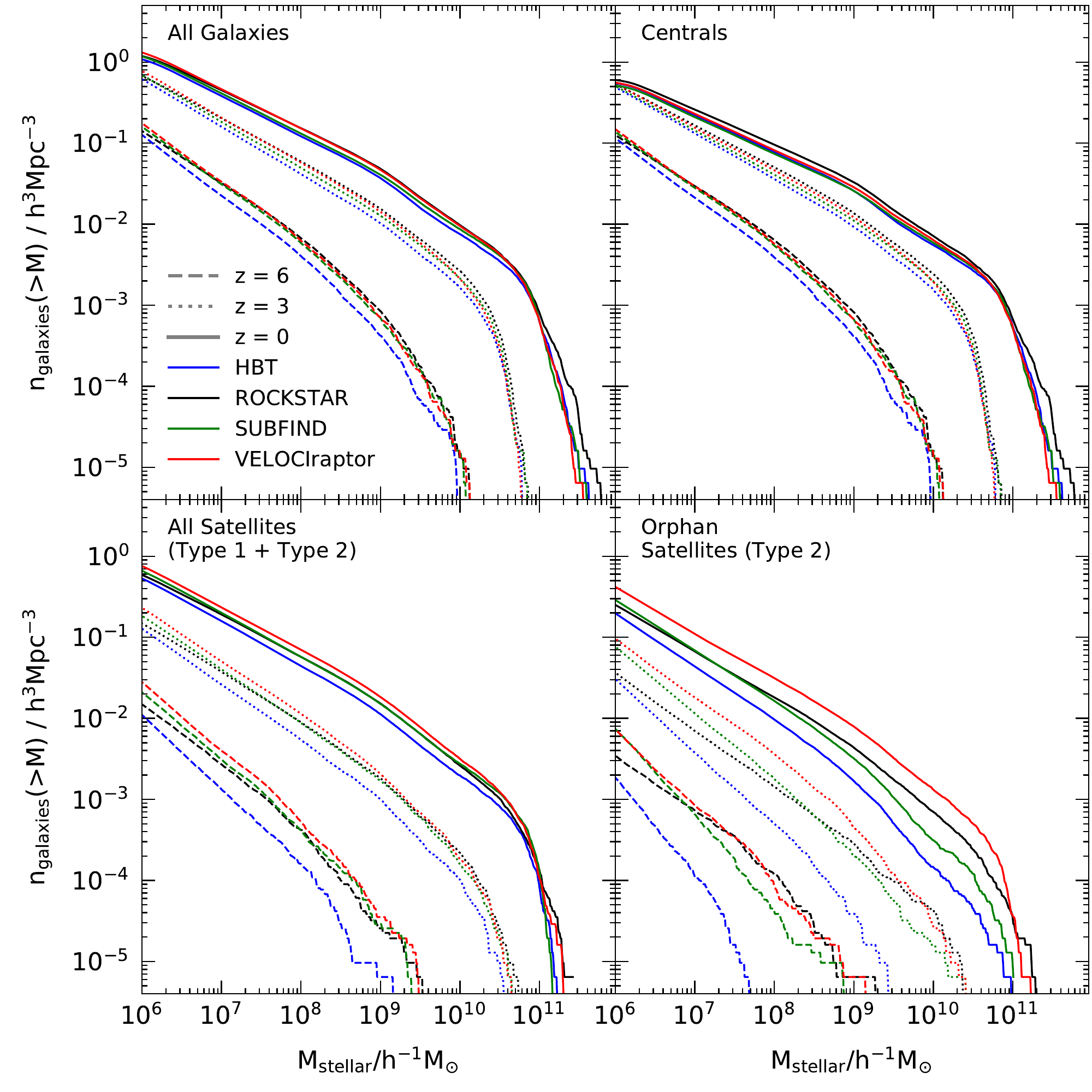}
    \caption{Cumulative stellar mass function for all galaxies (centrals + all satellites), centrals, all satellites (types 1 and 2), and type 2 satellites galaxies at redshifts $z=0$, $z=3$ and $z=6$ for the four halo finders, as labelled.}
    \label{fig: csmf}
    \end{figure*}
    
    \begin{figure}
    \centering
    \includegraphics[scale=0.55]{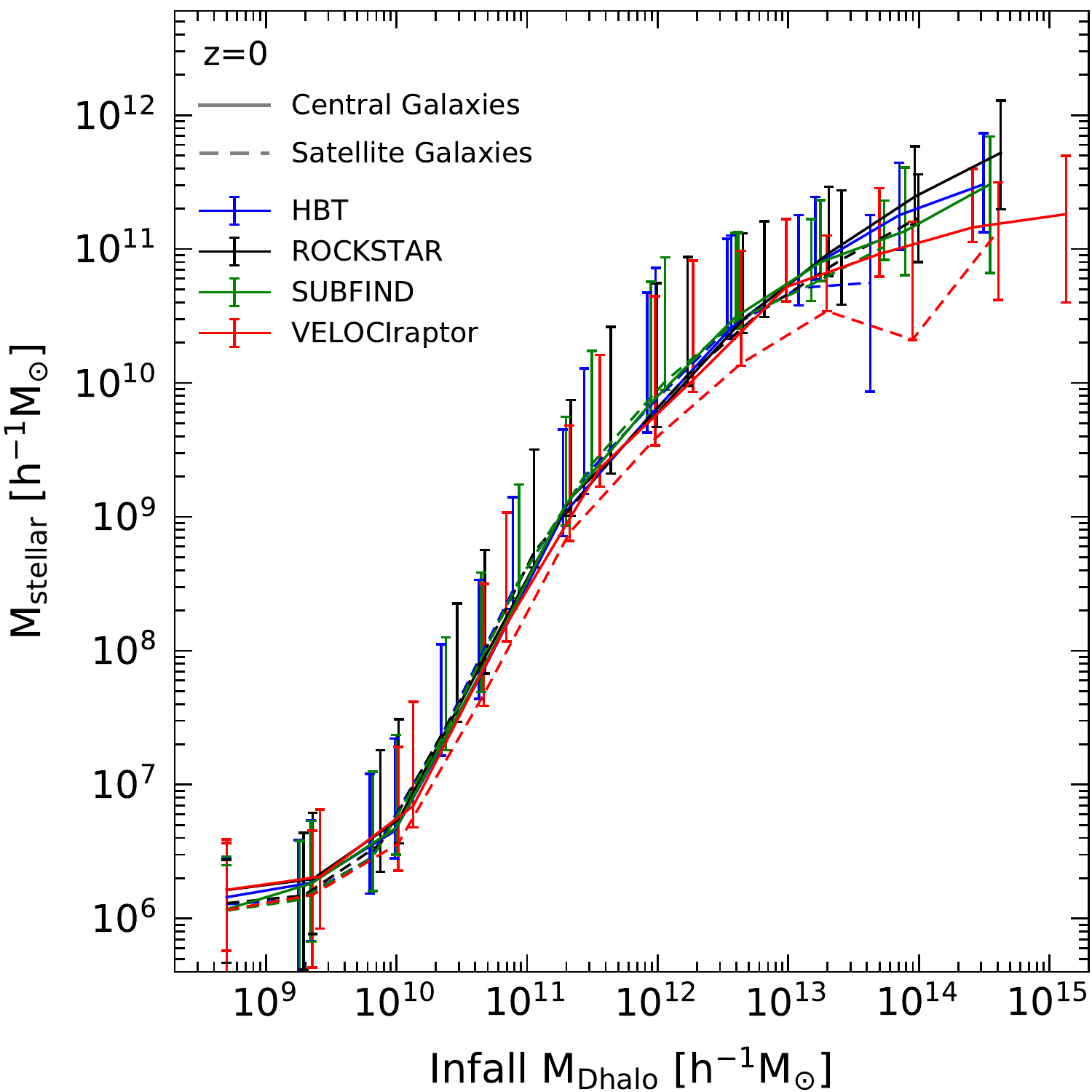}
    \caption{\galform~stellar vs. halo mass relation at $z=0$ for the different halo finders (different colours, indicated in the figure key). The halo mass used is the Dhalo mass for central galaxies (solid lines), and the Dhalo mass at infall for all satellite galaxies (dashed lines, including type 1 and type 2 satellite galaxies). The lines show the median and errorbars correspond to the 10 and 90 percentiles.}
    \label{fig:mstinfallmass}
    \end{figure}

    \begin{figure}
    \centering
    \includegraphics[scale=0.55]{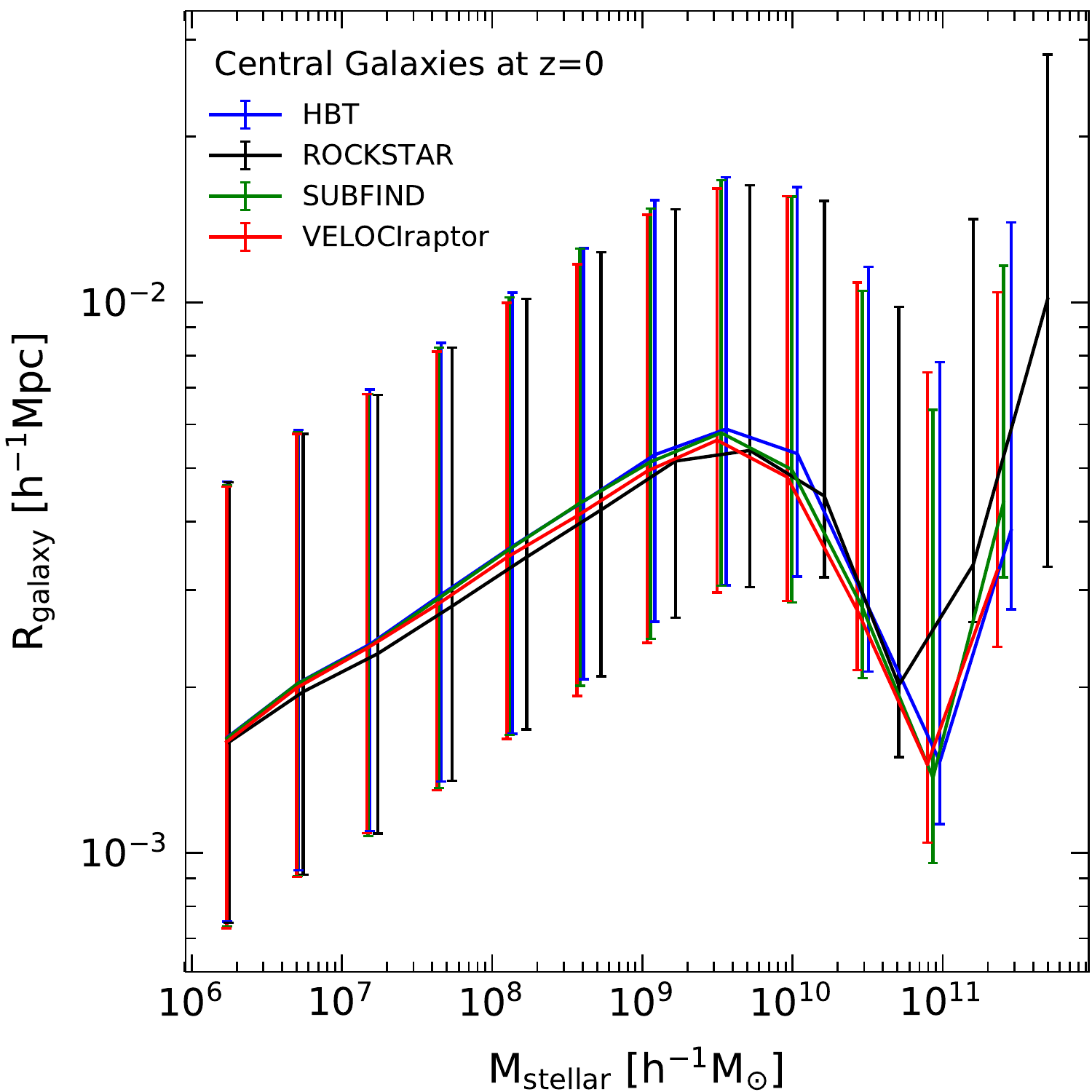}
    \includegraphics[scale=0.55]{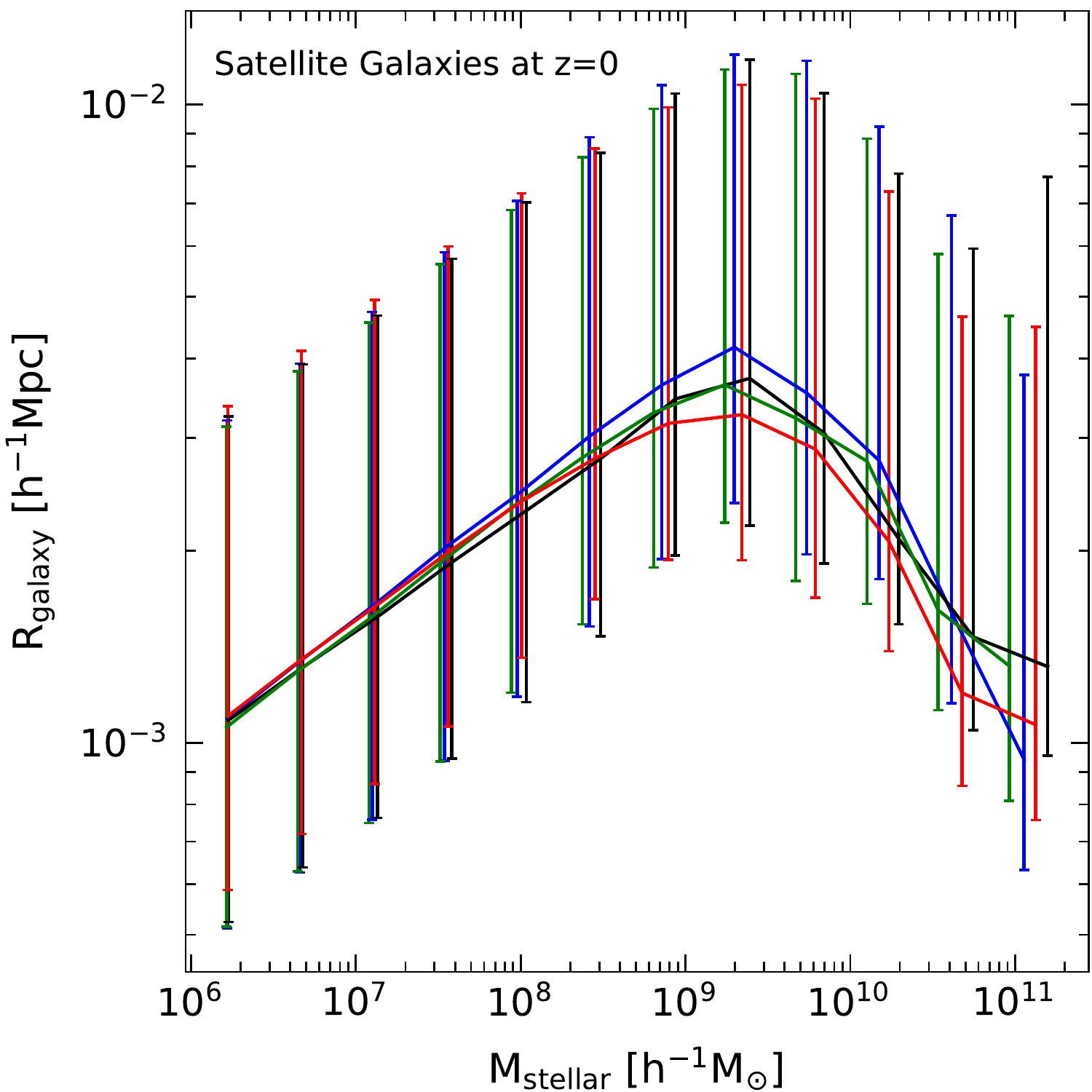}
    \caption{Median $r$-band half light radius against  stellar mass at $z=0$, for centrals (top) and satellites (both types; bottom). The bars show the {10 to 90 percentile} range.  Colours denote different halo finders (see  key).}
    \label{fig:radii}
    \end{figure}

    \begin{figure}
    \centering
    \includegraphics[scale=0.55]{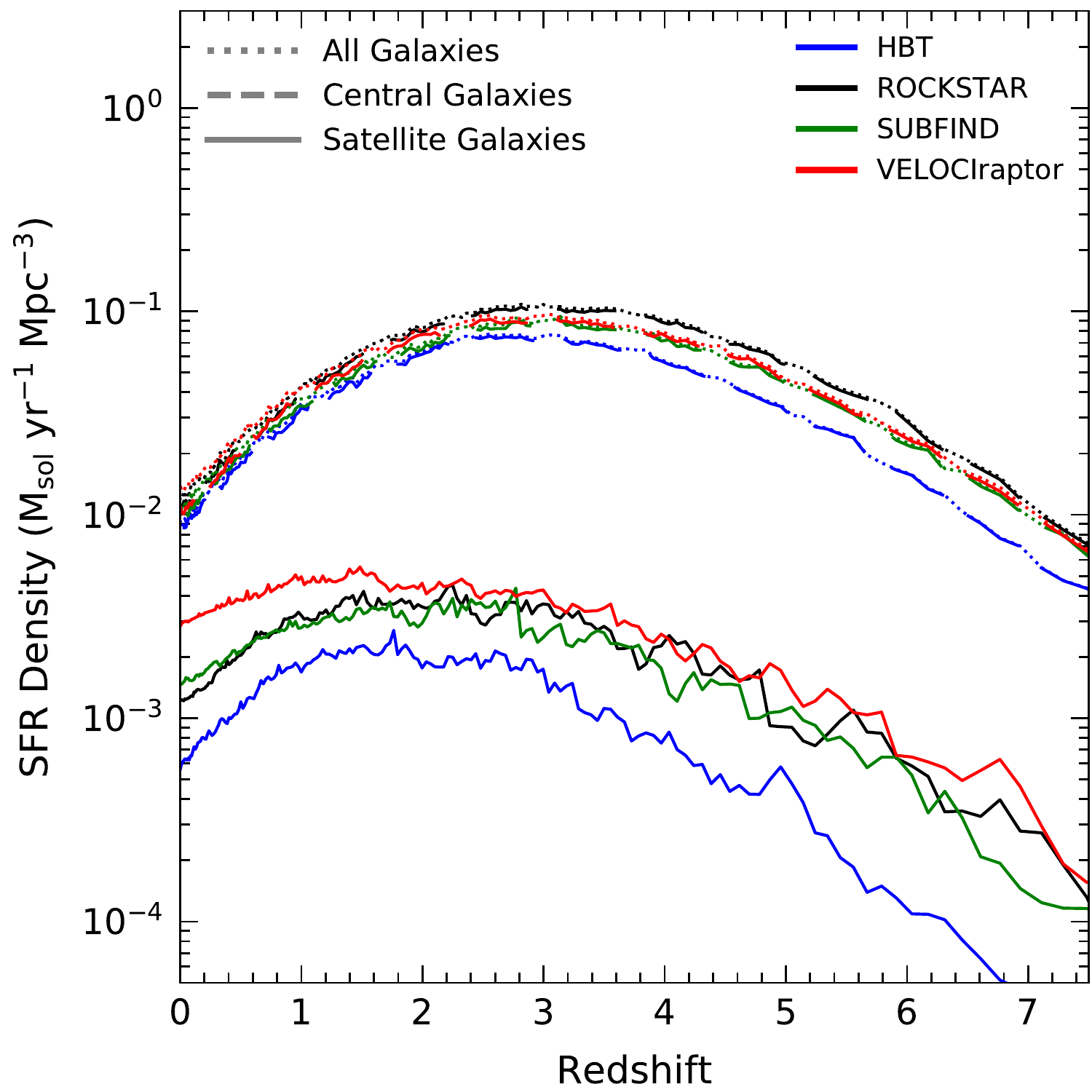}
    \caption{Star formation rate density as a function of redshift for \galform~galaxies resulting from the different halo finders. Dotted, dashed and solid lines show the results for all galaxies, centrals and satellites (type 1 and type 2), respectively.  {Note that the  lines for centrals are almost underneath the ones for all galaxies.}}
    \label{Fig:sfrd}
    \end{figure}

    \begin{figure*}
    \centering
    \includegraphics[scale=0.5]{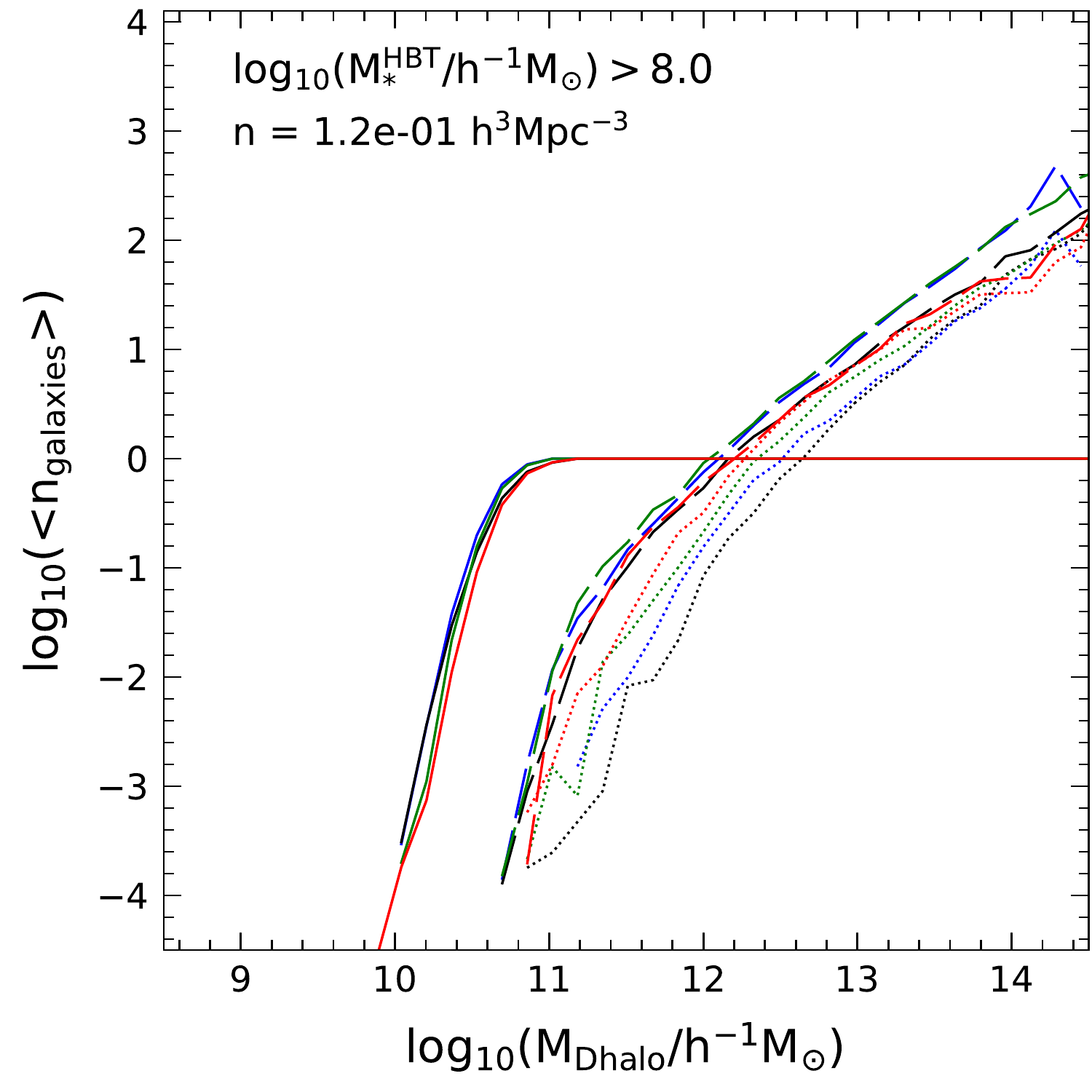}
    \hskip .3cm
    \includegraphics[scale=0.5]{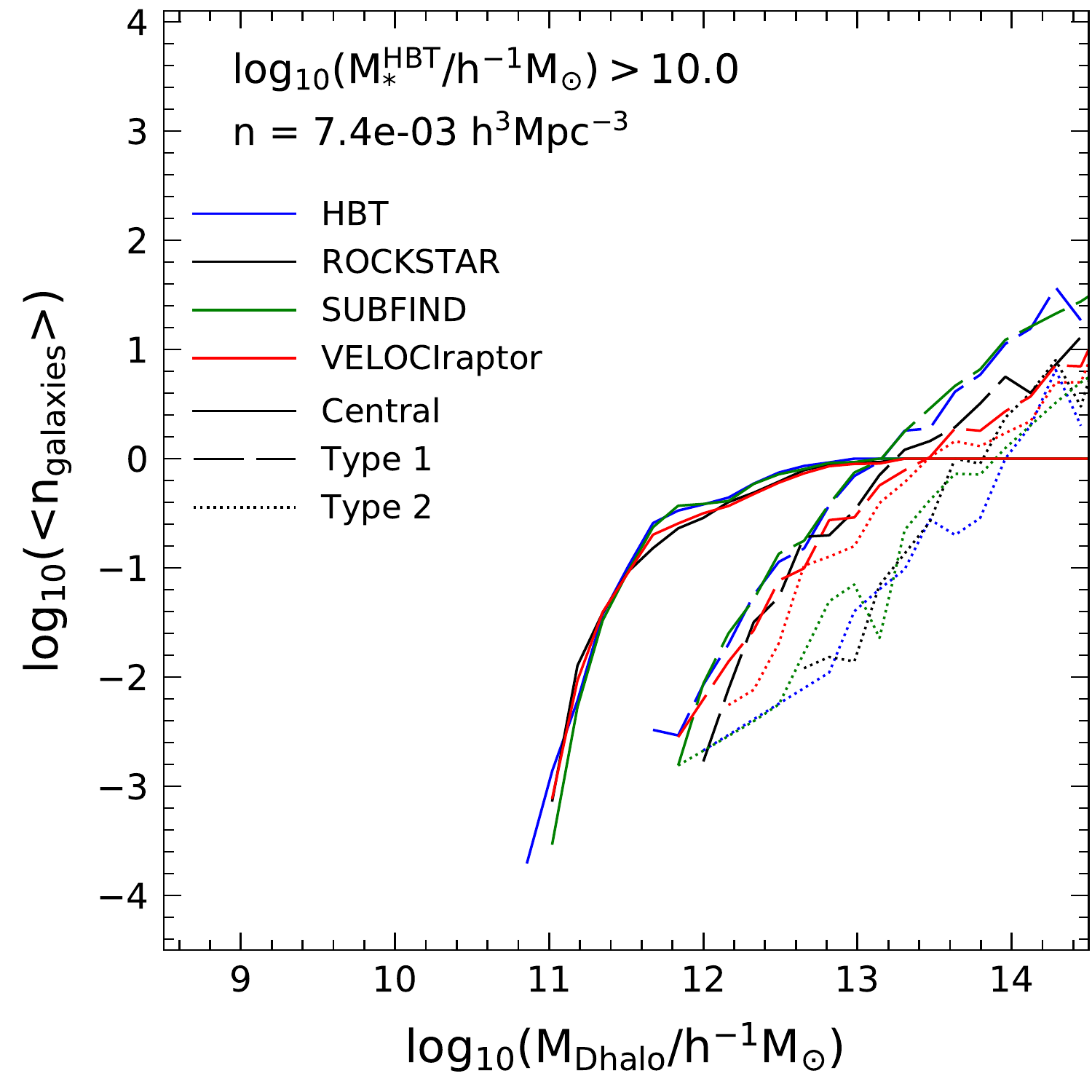}
    \vskip .3cm
    \includegraphics[scale=0.5]{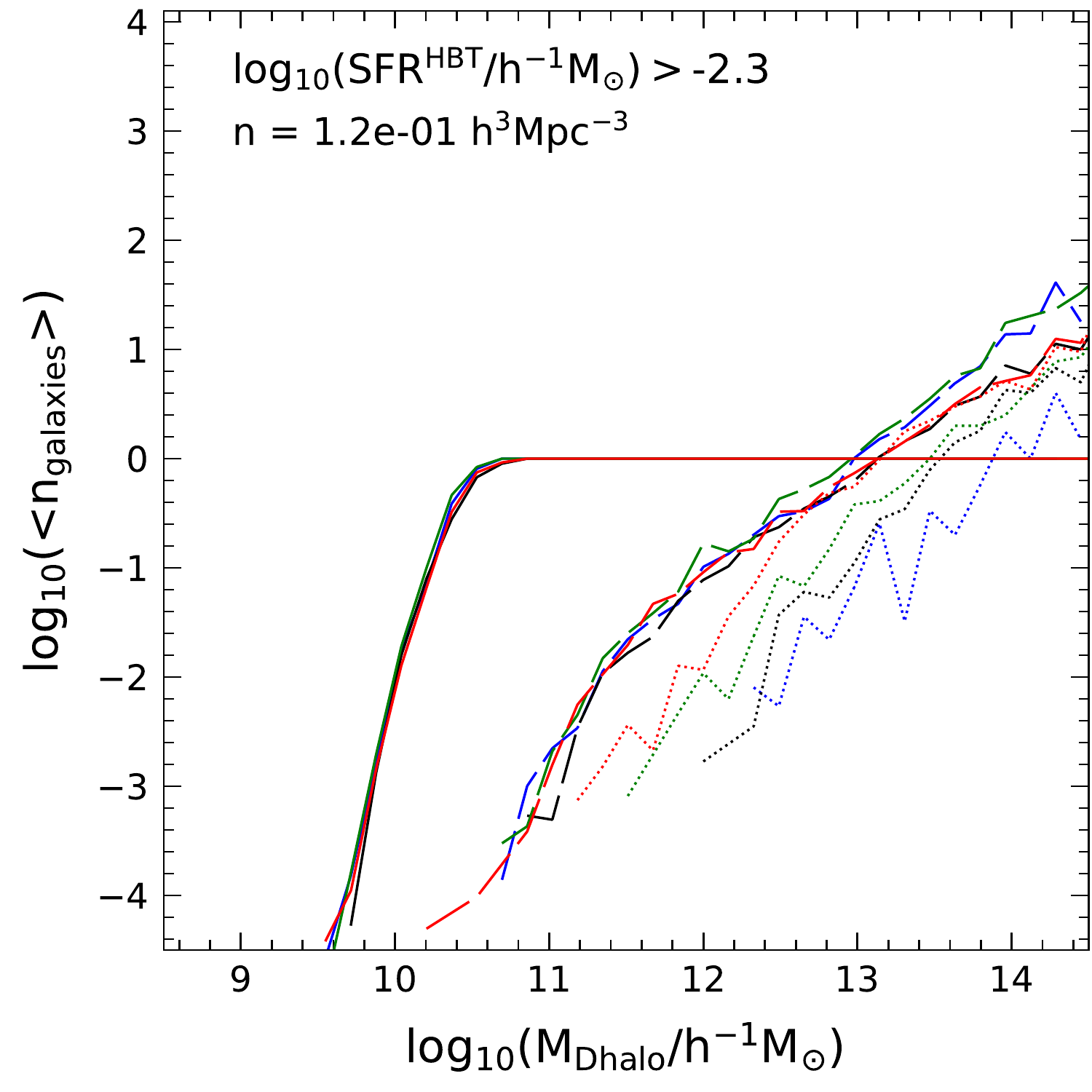}
    \hskip .3cm
    \includegraphics[scale=0.5]{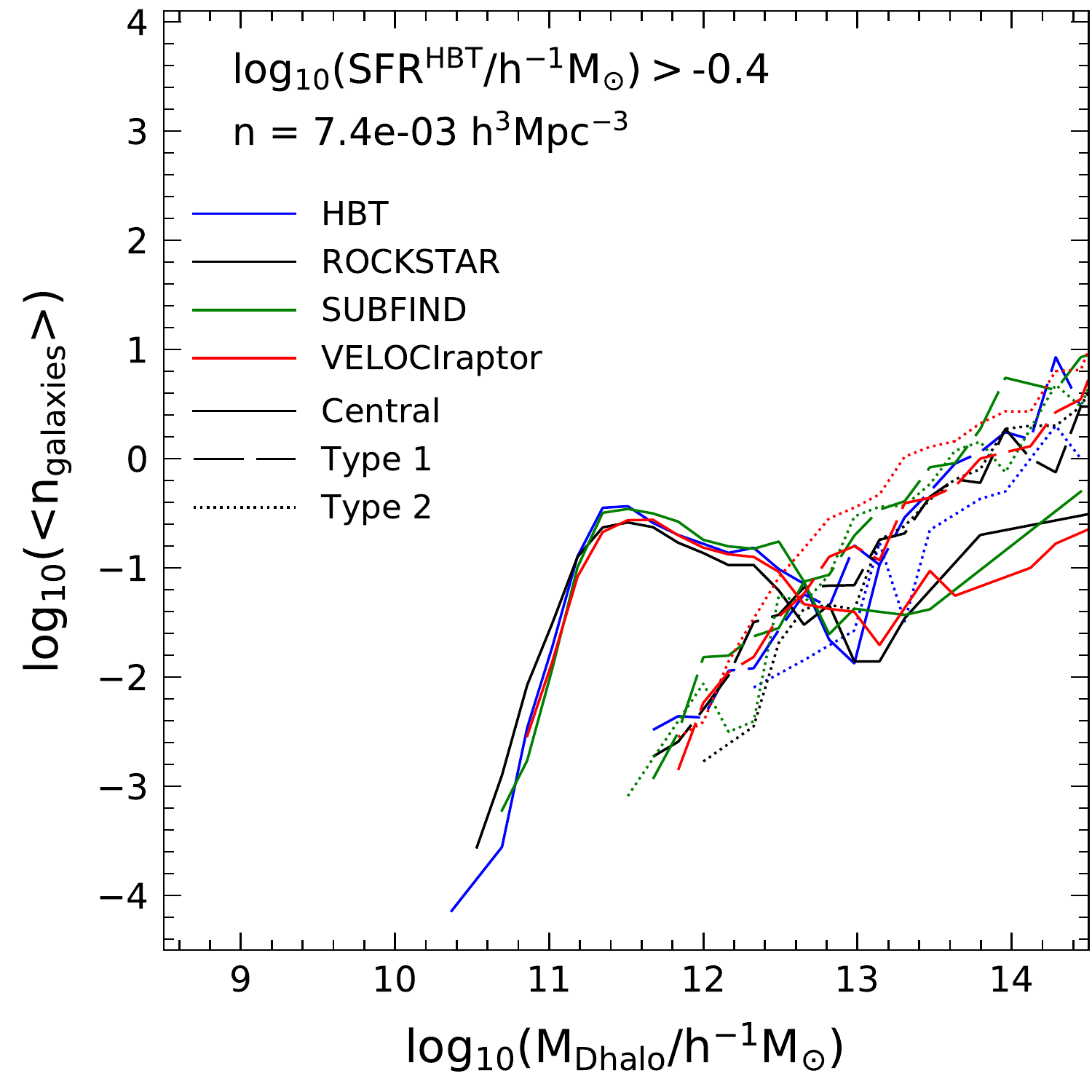}
    \caption{Halo occupation distributions of \galform~ samples run using the outputs of the four different finders (different colours as indicated in the figure) selected by stellar mass (top panels) and star formation rate (bottom panels) with space densities of $n=1.2e-1$h$^3$Mpc$^{-3}$ and $7.5e-3$h$^3$Mpc$^{-3}$ (left and right columns). The lines show the average number of galaxies as a function of the Dhalo mass, or infall mass for satellite subhaloes, for centrals, type 1 and type 2 satellites (different line types as indicated in the key).  The lower limits in stellar mass and SFR applied to define the sample are only shown for the \hbt~ run.}
    \label{panel:hod}
    \end{figure*}


{To better illustrate how in some cases satellites and Dhaloes can be identified differently by the different finders, we show some examples of the spatial distribution of Dhaloes and their satellites (and neighbour Dhaloes)  in } Fig.~\ref{fig: postion examples}. Each set of four panels corresponds to a {Dhalo matched between all finders} (following the procedure outlined in Section \ref{ssec:match}) and shows the positions of matched Dhaloes identified at $z=0$ (red circles) and {their} respective type 1 satellite subhaloes (blue circles). The radius of the circle plotted for each {Dhalo and type 1 satellite} is proportional to the logarithm of the Dhalo mass (at infall for type 1 satellite subhaloes), the dotted blue circles enclose the most distant {type} 1 satellite subhalo. { Yellow circles show neighbouring Dhalos.}

As expected, the {matched} Dhaloes are  centered on almost the same position. Several {type} 1 satellite subhaloes are identified by the four halo finders; when at least two finders detect them it can be seen that they show greater differences in their infall Dhalo masses than the Dhaloes themselves. \subfind~and \velociraptor~tend to find more {type} 1 satellites when processed through Dhaloes, as was already seen in Fig. \ref{fig: cdhmf}. The latter would combine with the ability of \velociraptor~ to detect subhaloes in higher density regions to produce the final differences in abundance of satellites among the different finders.  {Consequently, the finders that detect fewer satellites still detect some of these structures but in some cases as neighbouring Dhaloes (yellow).}

The population of satellites depends strongly on the abundances of Dhaloes {that contained them prior to infall to the current host Dhalo.} Fig.~\ref{fig: type 1 and 2 satellites subhaloes} shows the infall mass function for both {type} 1 (left panel) and {type} 2 (right panel) satellite subhaloes of matched Dhaloes. 
{The abundance of type 1 satellites is similar for those hosted by the highest mass Dhaloes. For the satellites hosted by lower mass Dhaloes, the abundance of \velociraptor\ subhaloes tends be higher than that of the other finders for high infall masses.  On the other hand,} there is an excess of type 2 satellite subhaloes in \velociraptor~compared with the other halo finders, regardless of the host Dhalo mass, although \rockstar~also shows some excess relative to \hbt~and \subfind~{for masses higher than $10^{10}$\Mh}; in the next {section} we test whether this is reflected in the number of \galform\ satellite galaxies.

The abundance of type 2 galaxies also depends on the dynamical friction timescales calculated for satellite galaxies once their host subhalo can no longer be resolved. In \galform, this timescale is assumed to depend on the mass ratio of the satellite subhalo to host Dhalo, as well as the orbital parameters of the satellite subhalo, at the time it becomes a type 2. The dynamical friction timescale is longer for smaller mass ratios, which could apply to more type 2 galaxies. 
In Fig.~\ref{fig: ratio sat1/dhalo} we plot this mass ratio, which is calculated using the satellite subhalo mass without any processing by Dhalos and \galform, divided by the Dhalo mass at the time of merger. 
For this plot, we identify mergers that occur between $z=0.01$ and $z=0$, for Dhaloes matched at $z=0$ in the four halo finders. 
{We select $z=0.01$ for the distributions to avoid} the last snapshot of the simulation as Dhalo needs future snapshots to clean its merger trees. The distributions are similar for the different finders, regardless of the Dhalo mass (we have also explored other redshifts and reach the same conclusion). {We therefore expect that \galform~will calculate very similar dynamical friction timescales for satellite galaxies once their host subhaloes change from type 1 to type 2, regardless of the halo finder used.} This makes us expect that the relative abundances of type 2 satellite galaxies will reflect the differences in {the} abundances of {type 2} satellite subhaloes  between the different halo finders  {(cf. Fig.~\ref{fig: cshmf})}.


\section{predicted galaxy properties for the different halo and merger tree finders}
\label{section: gal_galform}

We next study the effects of the different halo finders and merger tree builders on the properties of galaxies predicted by \galform. We look both for properties that are insensitive to the choice of merger tree builder and those which change. We run \galform~on the different merger trees keeping the model parameters fixed at the values selected in \citet{Baugh2019} for the \citet{Lacey2016} model. 

In order to focus on results that are not affected by the resolution limit of the simulation, we first run \galform~on the output of the four halo finders using two different lower limits on {subhalo} mass,{ applied before the monotonicity condition is imposed}. The first cut corresponds to $40$ dark matter particles or $3.12\times 10^{8}$\Mh, and the second to $400$ particles, i.e. $3.12\times 10^{9}$\Mh.  We {measured} the stellar mass functions for the eight runs and looked at what stellar mass the cumulative stellar mass functions of runs with different lower subhalo mass limits start to diverge from one another; this happens around a stellar mass of $10^7$\Mh~for all finders, which we interpret as the resolution limit for the runs using subhaloes of $400$ or more dark matter particles. Based on this, we {conservatively} estimate that the resolution limit in stellar mass should be around $10^6$ \Mh~or lower for a halo mass resolution limit of 40 particles, as used in our standard Dhalo catalogues. {Therefore} from this point forward we only use galaxies with stellar masses $\geq 10^6$\Mh.

\subsection{Galaxy Stellar masses}

We start the comparison of model outputs with the different finders with Fig.~\ref{fig: csmf}, which shows the cumulative stellar mass function.

The number of central galaxies depends on the number of Dhaloes available to host them. For \rockstar, \subfind~ and \hbt, the comparison of the stellar mass function of central galaxies at $z=0$ shown in Fig.~\ref{fig: csmf} is similar to that for the cumulative mass function of Dhaloes at $z=0$ shown Fig.~\ref{fig: cdhmf}; \rockstar~has more Dhaloes and central galaxies than the other two halo finders over almost the entire mass range.  Central galaxies from the \velociraptor\ run do not show the excess seen for central Dhaloes in Fig.~\ref{fig: cdhmf} and are consistent with the abundances of galaxies from the \hbt~ and \subfind~ runs.

As {type 1 satellite} galaxies are hosted by {resolved} subhaloes, their number density is directly related to the number of {type 1} satellite subhaloes, especially in the \citet{Baugh2019} version of \galform~that only allows galaxy mergers after their host satellite subhalo has been lost. Figs.~\ref{fig: cdhmf} and \ref{fig: csmf} show general consistency between the relative abundances of $z=0$ satellite subhaloes and satellite galaxies for the different finders, although the differences are smaller in the stellar mass functions. We find {that the} larger number of satellite subhaloes for \subfind~and \velociraptor~ corresponds to larger numbers of satellite galaxies with these halo finders. 

Type 2 galaxies are indicative of satellite subhaloes that have been lost.  The stellar mass functions show more {type} 2 galaxies with \velociraptor~and \rockstar~{than \hbt~and \subfind}. As mentioned above, this excess of type 2 galaxies is related to the number of  {satellite subhaloes that merged with the central subhalo.
This shows a consistent picture involving merged subhalo progenitors 
and the stellar mass functions of type 2 satellites as there are also larger numbers of type 2 satellite subhaloes with \velociraptor~and \rockstar~ than with the other two finders. }
  Fig.~\ref{fig: cdhmf} (right) shows the Dhalo mass function of {type 2 satellite} subhaloes. Here the numbers are the lowest for \hbt, as is the case for type 2 galaxies, followed by \subfind, \rockstar~ and \velociraptor~in increasing order, which roughly matches the relative numbers of type 2 galaxies. The number of type 2 galaxies is further shaped by the dynamical friction timescale that elapses before a galaxy merges with the central galaxy of the Dhalo.  Fig.~\ref{fig: ratio sat1/dhalo} shows the distribution of ratios of satellite subhalo to Dhalo masses for merging satellites. It can be seen that there are practically no differences between the finders. Thus the number of type 2 subhalos is the main driver of the relative abundances of type 2 galaxies {for different halo finders}.


{In summary}, as the subhalo definition depends on the finder, the abundance of the different galaxy types depends on the tree builder. \rockstar~produces slightly more high mass galaxies than the other finders because it yields more high mass central subhaloes. Although \hbt~produces more central galaxies than \subfind, and \subfind~produces more type 1 satellite galaxies than \hbt, these two halo finders produce similar numbers of galaxies overall since there is a similar total number of subhaloes at each mass as shown in Section~\ref{section: Merger Trees}. \velociraptor~ produces more satellite galaxies than the other finders due to the combination of a higher abundance of satellite subhaloes with a larger population of type 2 satellites.

\subsection{Comparison of other galaxy properties}

We now focus on the relation between stellar mass and halo mass, galaxy sizes and the evolution of the star formation rate density.
  
The efficiency {of} star formation in a halo, measured by $M_{\rm stellar}/M_{\rm halo}$, is mostly set by the assumptions in the galaxy formation model \citep{Mitchell2016}, so we do not expect this to vary significantly with the halo finder. This is confirmed in Fig.~\ref{fig:mstinfallmass}, where we show the relation between stellar mass and Dhalo mass at redshift $z=0$ for centrals and satellites. 
For centrals (solid lines) the infall mass is simply the Dhalo mass, whereas for satellite galaxies (dashed)  the infall mass corresponds to the Dhalo mass before the subhalo became a satellite. The relations are mostly indistinguishable between the different finders, but there is a slight tendency for \velociraptor~galaxies to show lower stellar masses at fixed infall Dhalo mass for Dhalo masses {above} $\sim 10^{13}$\Mh~for centrals and at all Dhalo masses for satellites. This could be due to the same galaxies having higher Dhalo masses in \velociraptor~compared to the other halo finders, as shown in Fig. \ref{Fig: percentages mixed}.

Another important property that could be affected by the merger trees {is} galaxy sizes,  {as mergers can induce bursts of star formation and thus regulate the relative amount of stars in the spheroid and disc components of a galaxy}. Fig.~\ref{fig:radii} shows the $r$-band half light galaxy radius (as defined in \citet{Lacey2016}). 
Central galaxy sizes are very similar for \hbt, \rockstar~and \subfind, with \rockstar~showing $\sim10$ per cent smaller sizes {over} almost the entire stellar mass range, except for {very} high masses. For satellite galaxies the differences in the mean sizes are larger, {with \hbt~showing larger sizes over almost the entire stellar mass range, except for very high masses where the sizes are lower than the other halo finders. } \velociraptor~shows smaller sizes for stellar masses $>10^9$\Mh. A smaller size is in general related to a larger stellar mass for the spheroid component, which in turn can be due to an earlier or more rapid star formation history. By looking at the evolution of star formation in galaxies we can clarify these differences.

Fig.~\ref{Fig:sfrd} shows the star formation rate density (SFRD) for \galform~galaxies as a function of redshift for the different halo finders. \rockstar~trees give a slightly higher SFRD for central galaxies at all redshifts. {\hbt~shows a lower SFRD than the other halo finders  at all redshifts for central and satellite galaxies}, which is explained {by} the lower cumulative Dhalo mass function at all redshifts in \hbt~(see Fig.~\ref{fig: cdhmf}) impacting the cumulative stellar mass function at all redshifts (see Fig.~\ref{fig: csmf}). A higher SFRD could be related to a larger spheroid component, and smaller galaxy size, which makes the higher SFRD and the smaller sizes of satellites in the \velociraptor~run consistent in this simplified picture (cf. Fig.~\ref{fig:radii}).

\subsection{Halo occupation distributions}

A key objective of galaxy formation models is to connect the cosmological model with the observed clustering of galaxies. Here, instead of measuring the spatial correlation function of \galform~ galaxies resulting from the different finders we will look at their halo occupation distribution, that is, the  mean number of galaxies as a function of halo mass, as this metric is directly related to their clustering \citep{Berlind02,Zheng05}.  

We consider two samples for this end, one selected with a lower limit on stellar mass, the other with a lower limit on star formation rate.  The first aims at producing samples similar to those obtained by selecting target galaxies {based on broadband} optical luminosity (e.g. SDSS legacy, eBOSS LRGs, LSST, DESI BGS), whereas the second approximates selection by emission line luminosity (such as the emission line galaxy samples from eBOSS, Euclid and DESI).  

We consider samples with space densities of $n=1.2\times10^{-1}h^{3}$Mpc$^{-3}$ and $7.5\times10^{-3}h^{3}$Mpc$^{-3}$, which, for reference, correspond to applying stellar masses cuts of at least $10^{8}$\Mh~and $10^{10}$\Mh, respectively, to the \galform~run using the \hbt~halo finder.  Note
that because the EAGLE100 simulation has a comparatively small volume, the space densities that we can reliably probe are somewhat higher than those expected for the samples mapped by current and future surveys, but the results we find here are still valid for the comparison we perform  between different finders.

Fig.~\ref{panel:hod} shows the HODs for the stellar mass and SFR selected samples, for high and low space densities.  The plot shows the values for the cuts applied on stellar mass and SFR for \hbt~ only, as these values are slightly different {for the other halo finders}.  Predictions are shown for centrals, and type 1 and 2 satellites. The results are compatible with what has been presented in the previous subsections, and show that the central occupation in the high density samples of SFR and stellar mass selected samples are almost indistinguishable between the outputs of the different finders.  The Dhalo mass at which the central occupancy reaches $1$ differs by less than 0.05 dex.  The occupancy of the type 1 satellites is also very similar between the finders (there are some differences at higher halo masses). \velociraptor~does not appear to have more type 1  satellite galaxies than the other halo finders but shows a higher number of type 2 satellite galaxies (cf. Fig.~\ref{fig: type 1 and 2 satellites subhaloes}). Only the type 2 {satellite galaxies} show, albeit with higher noise, a higher occupancy for \velociraptor~ in both the SFR and stellar mass selections, with an excess with respect to the other finders that is similar to that seen in type 2 subhalos in Fig.~\ref{fig: cdhmf}. For the lower  number density samples the results are slightly noisier but the conclusions are the same as for the higher density samples.

The HOD for the low density SFR sample shows a central occupation that increases with mass, reaches a peak below unity, then drops before rising once more. This is due to the effect of AGN feedback  inhibiting quiescent SFR in massive haloes, but allowing some bursty star formation to take place. This effect, combined with the cut on SFR that defines the sample, is responsible for the shape of the central HOD, and is consistent with the literature (e.g. \citealt{Contreras2019}).  Even where the occupancy shows  complicated behaviour, the four halo finders give similar results.

We conclude that the use of different halo finders, processed through \galform, including the Dhalo preprocessing, produces samples with essentially the same {halo occupation implying that they should also show similar clustering.}

\subsection{Comparison between matched galaxies}

\begin{figure*}
    \centering
    \hspace{-0.1cm} \includegraphics[scale=0.41]{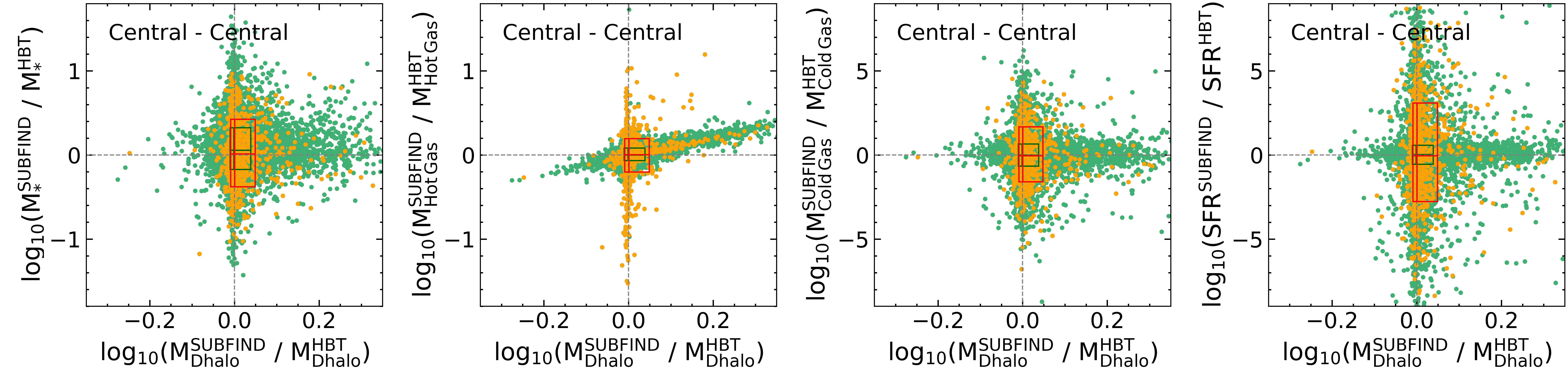}
    \vskip -.2cm
    \caption{Variation of properties of individual, matched galaxies 
    showing ratios of \mstellar~(left), \mhotgas~(second), \mcoldgas~(third) and \inststr~(right) as a function of the ratio of Dhalo mass M$_{\mathrm{Dhalo}}$ at $z=0$ for matched subhaloes in \hbt~and \subfind. Here we compare only galaxies that are central galaxies in both halo finders.  Light green dots show variations for low stellar masses $10^8<$ M$_*$/\Mh~$<10^{10}$ and orange dots correspond to high stellar masses M$_*$/\Mh~$>10^{10}$. The boxes delimit the 10 and 90 percentiles of the distribution of each axis for low and high stellar masses (green and red, respectively).}
    \label{panel: delta h1 r1 s1}
    \end{figure*}

We now look more closely at the variation in galaxy properties resulting from the use of different halo finders. Fig.~\ref{panel: delta h1 r1 s1} shows {a} scatter plot of the ratios of properties of central galaxies as a function  of the ratio of Dhalo mass between \subfind~ and \hbt~ for matched central subhaloes at $z=0$. 
For matched central {subhaloes} the Dhalo masses in these two finders are quite similar, with a small scatter (note that the scale on the x-axis is much smaller than on the y-axes) skewed to lower Dhalo masses for \hbt~ (the boxes extend further to the right). {The} offsets in the medians of the stellar, hot gas, cold gas masses, and even in star formation rates that result from running \galform~ on either finder are only a few per cent. 
The scatter plot shows that individual differences can be quite large, up to a factor of $10$ for stellar and hot gas masses, and as large as $10^5$ or even more {for} cold gas mass or SFRs.  However, $80$ per cent of the population of galaxies in either finder have stellar masses that agree within $\sim 20$ per cent, hot gas masses to $<5$ per cent, and cold gas masses and SFRs within a factor of $10$, increasing only slightly for the high stellar mass range. Note that the percentiles do not vary significantly between the two stellar mass ranges, except for SFRs for which the percentiles are narrower for low stellar mass than for high stellar mass.

    \begin{figure*}
    \includegraphics[scale=0.42]{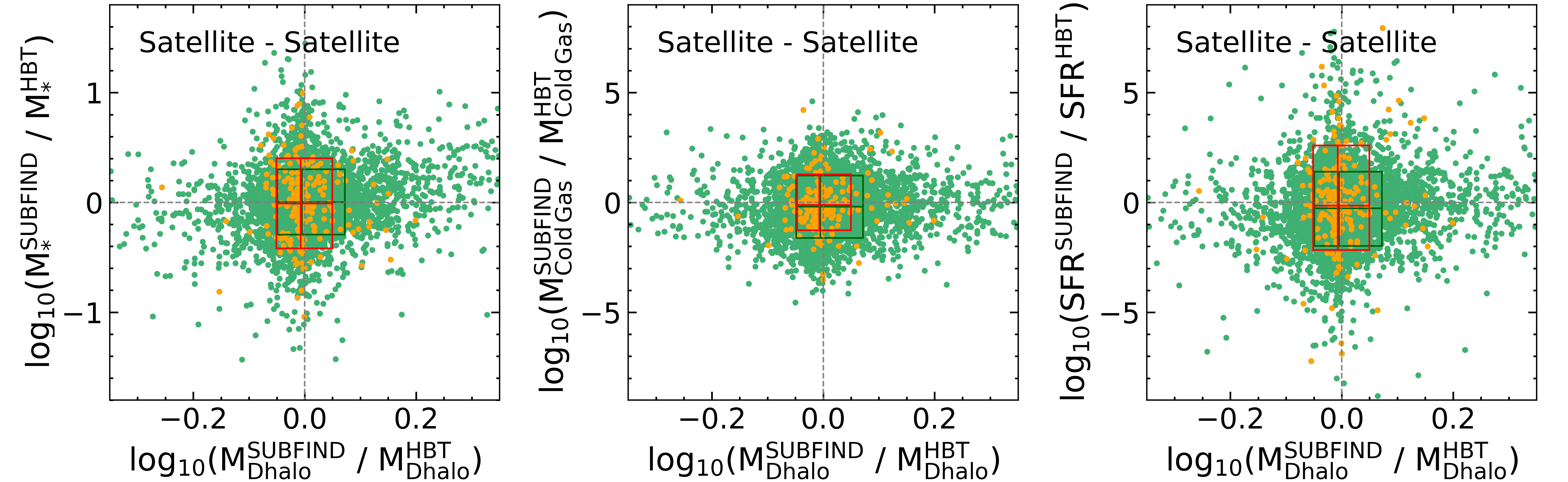}\\ \vspace{0.5cm}
    \includegraphics[scale=0.42]{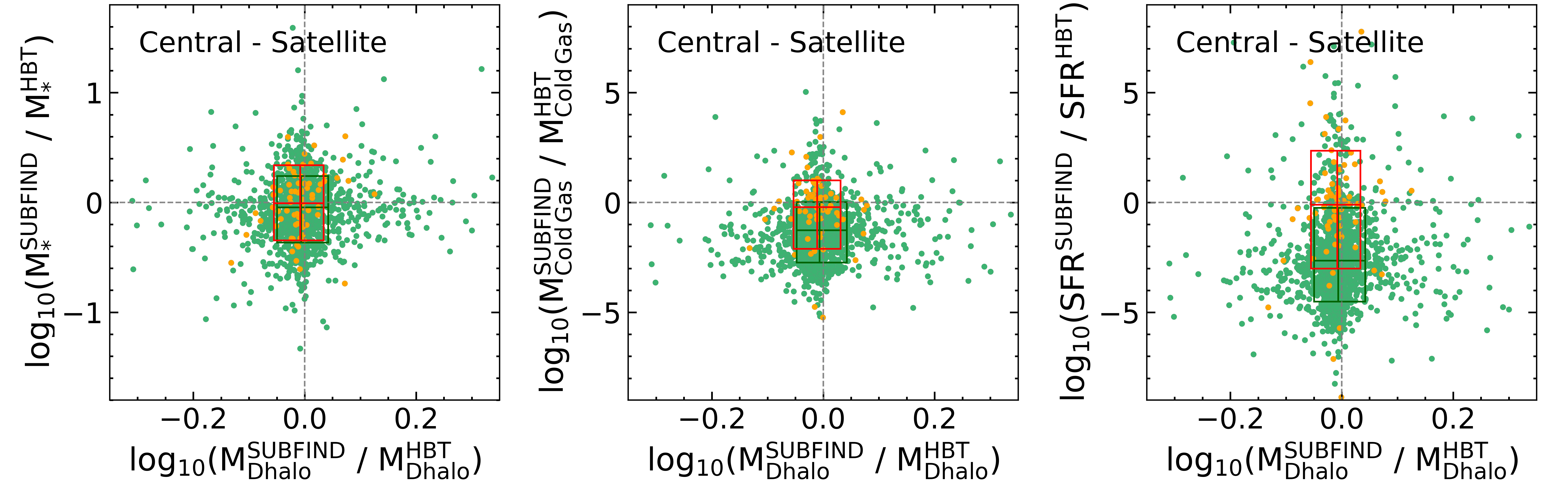}
    \caption{
    Same as Fig.~\ref{panel: delta h1 r1 s1} except for matched \hbt~and \subfind~satellite galaxies (top row), and \hbt~central galaxies matched to \subfind~ satellite galaxies (bottom row). In this figure the hot gas mass is not shown as it is zero by definition for satellites in this version of \galform. 
    }
    \label{fig:satdelta}
    \end{figure*}

    \begin{figure*}
    \centering
    \includegraphics[scale=0.28]{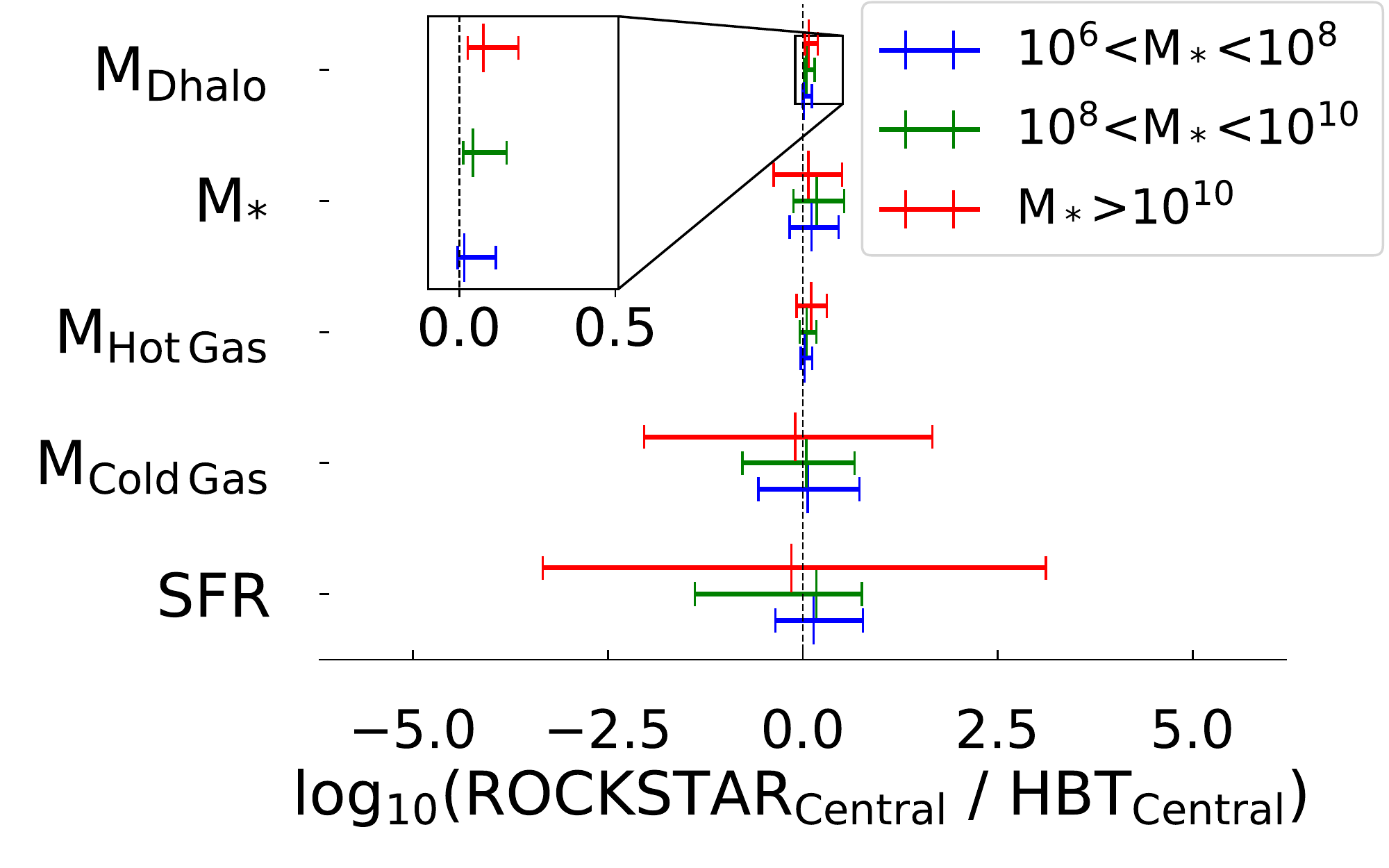} \includegraphics[scale=0.28]{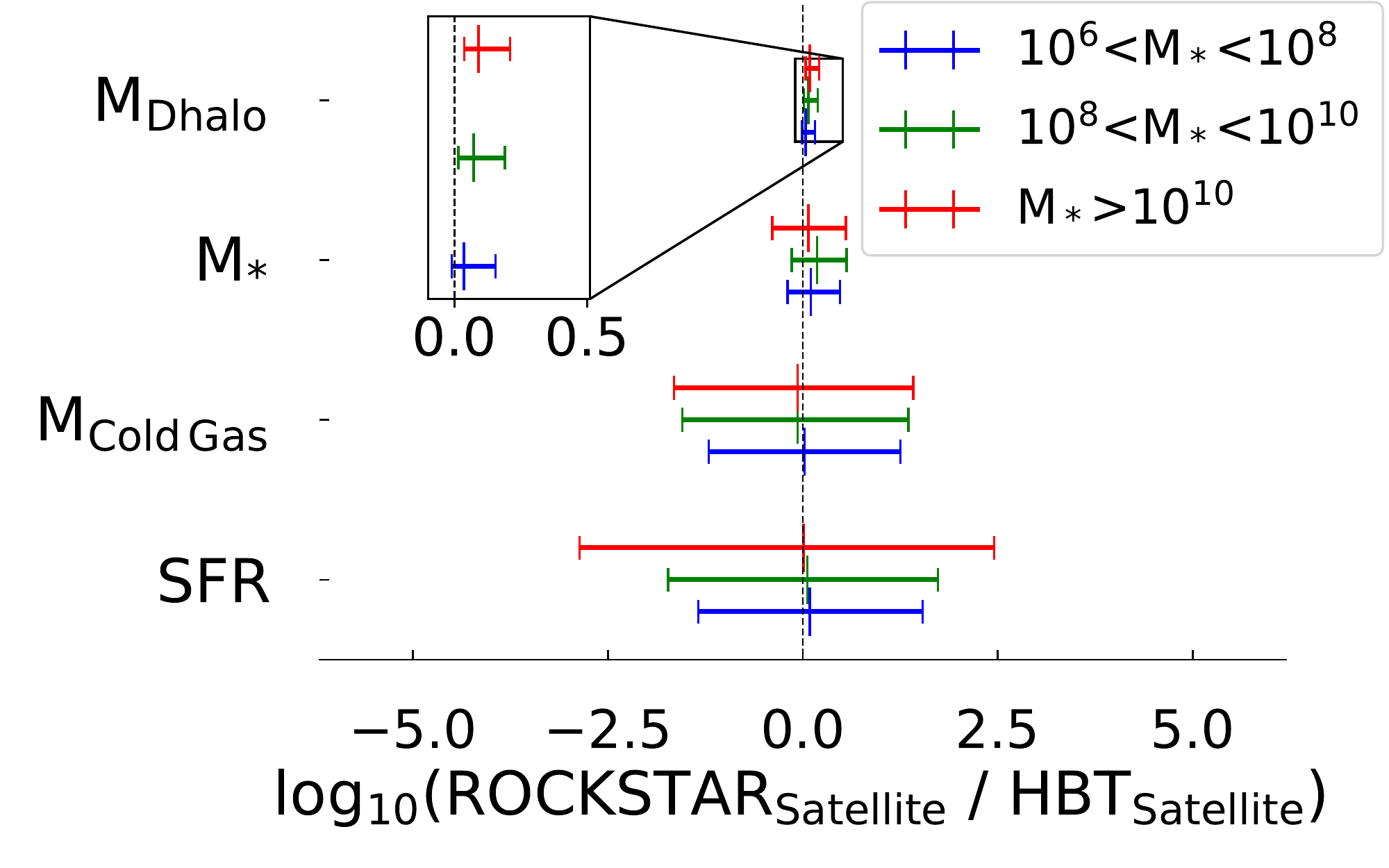}   
    \includegraphics[scale=0.28]{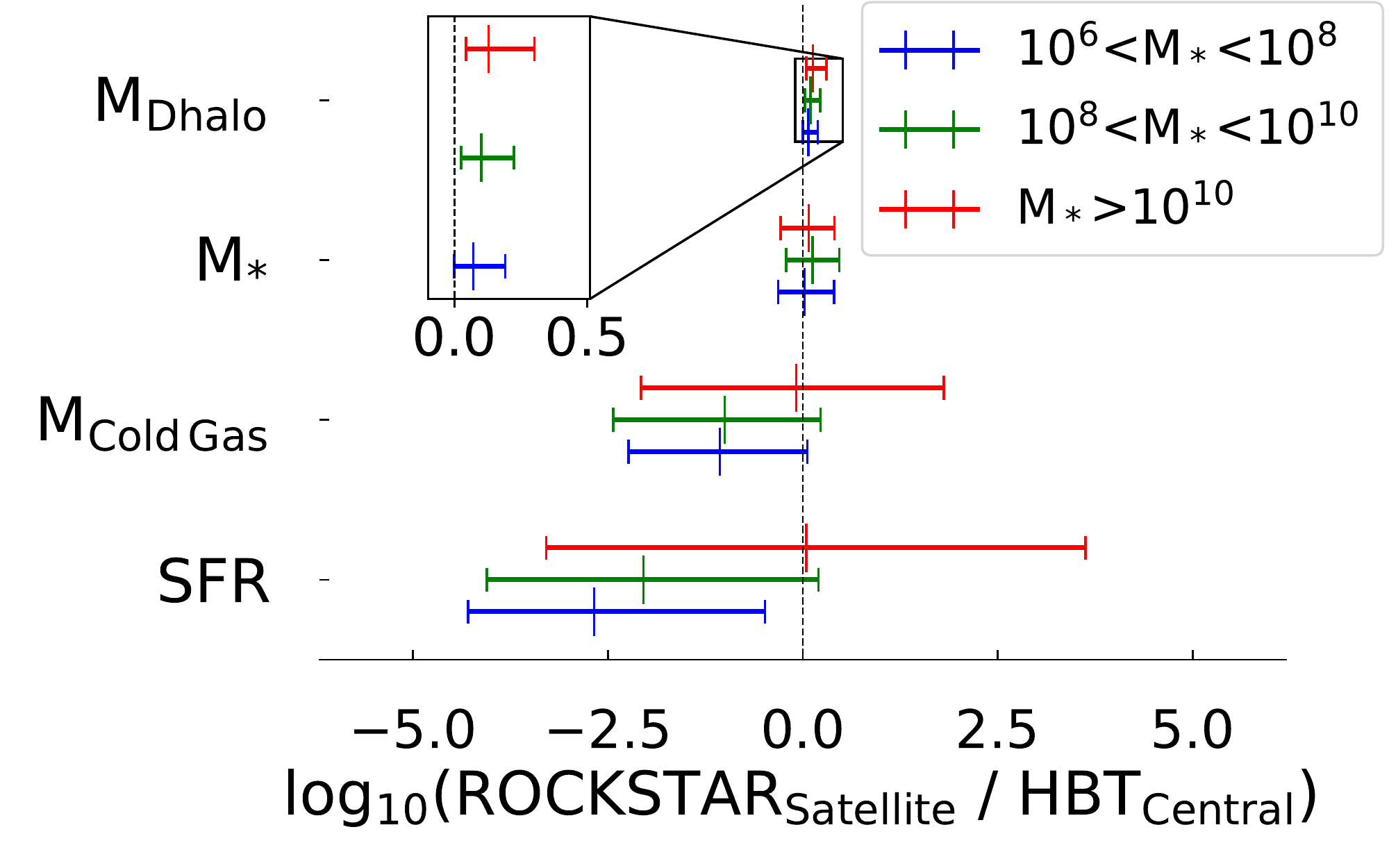}   \\   
    \vskip 0.5cm
    \includegraphics[scale=0.28]{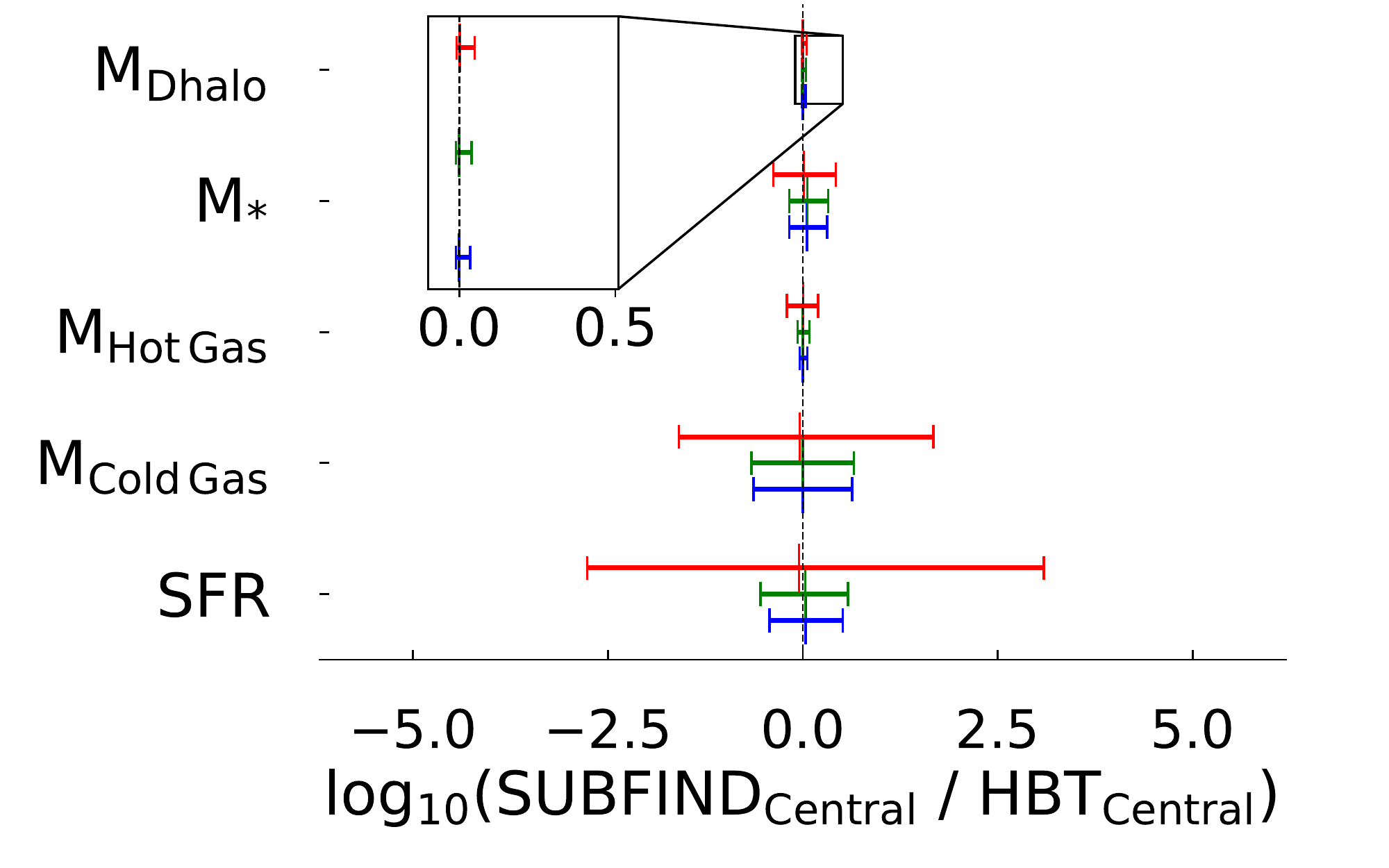}    
    \includegraphics[scale=0.28]{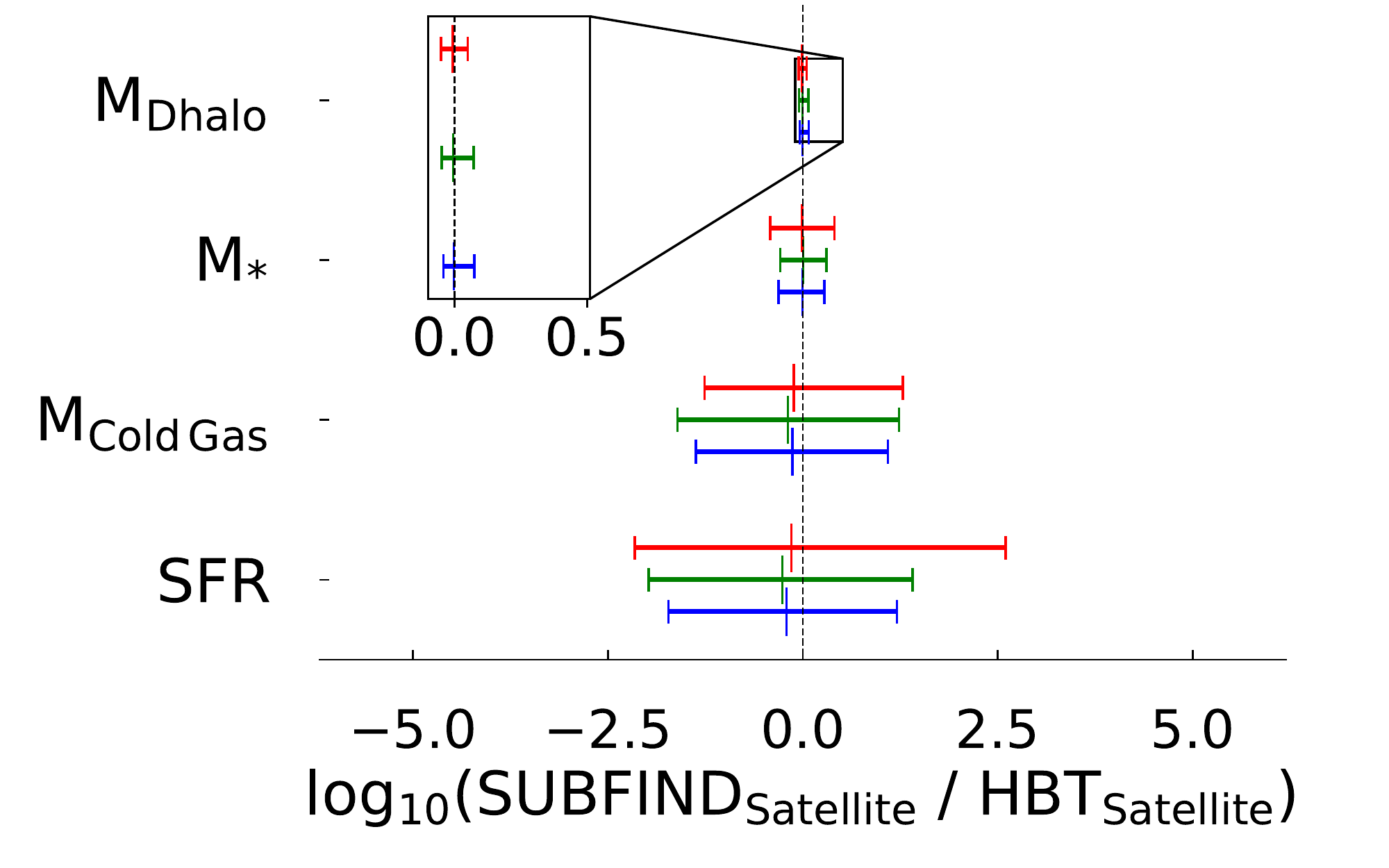}  
    \includegraphics[scale=0.28]{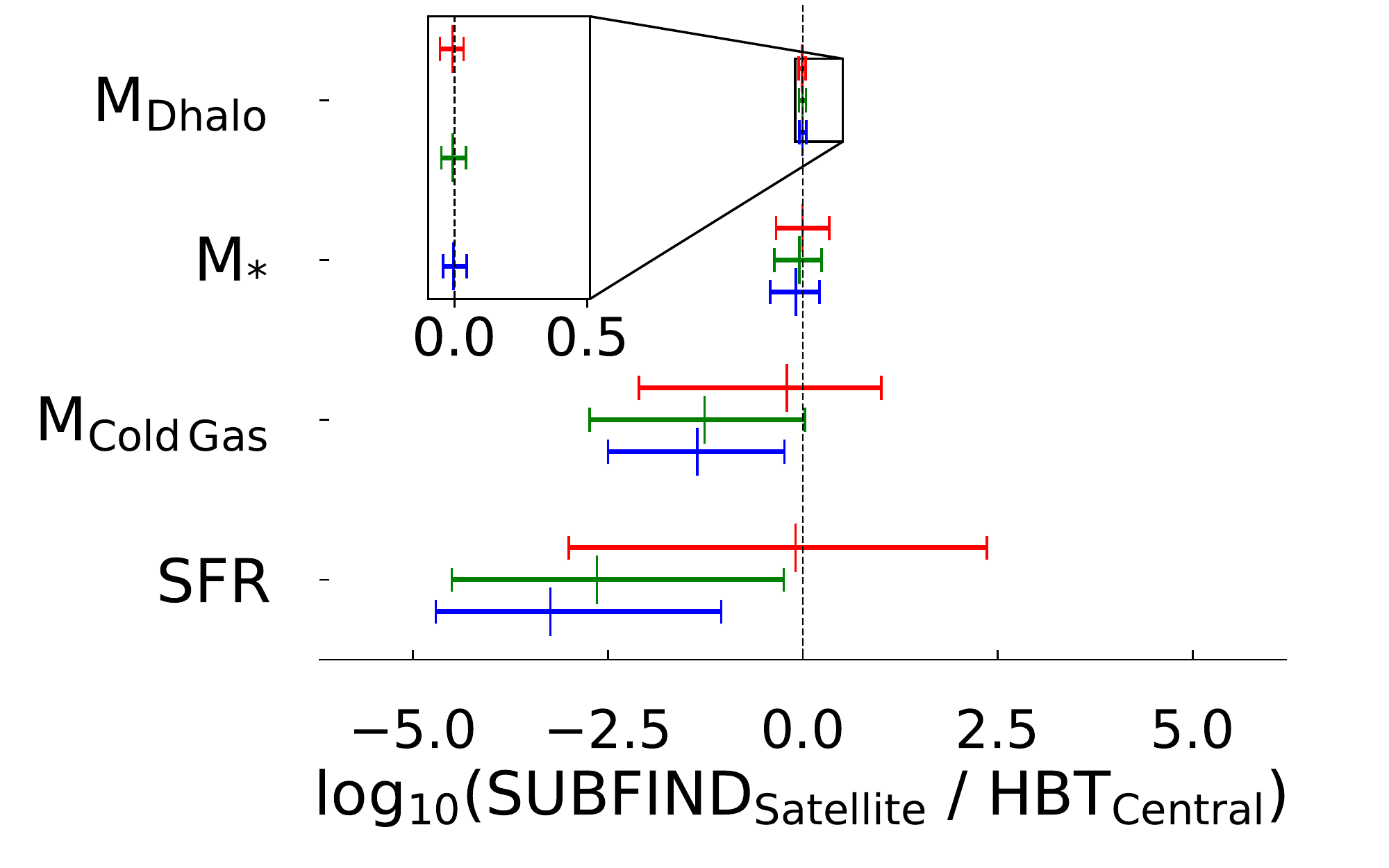}    \\
    \vskip 0.5cm
    \includegraphics[scale=0.28]{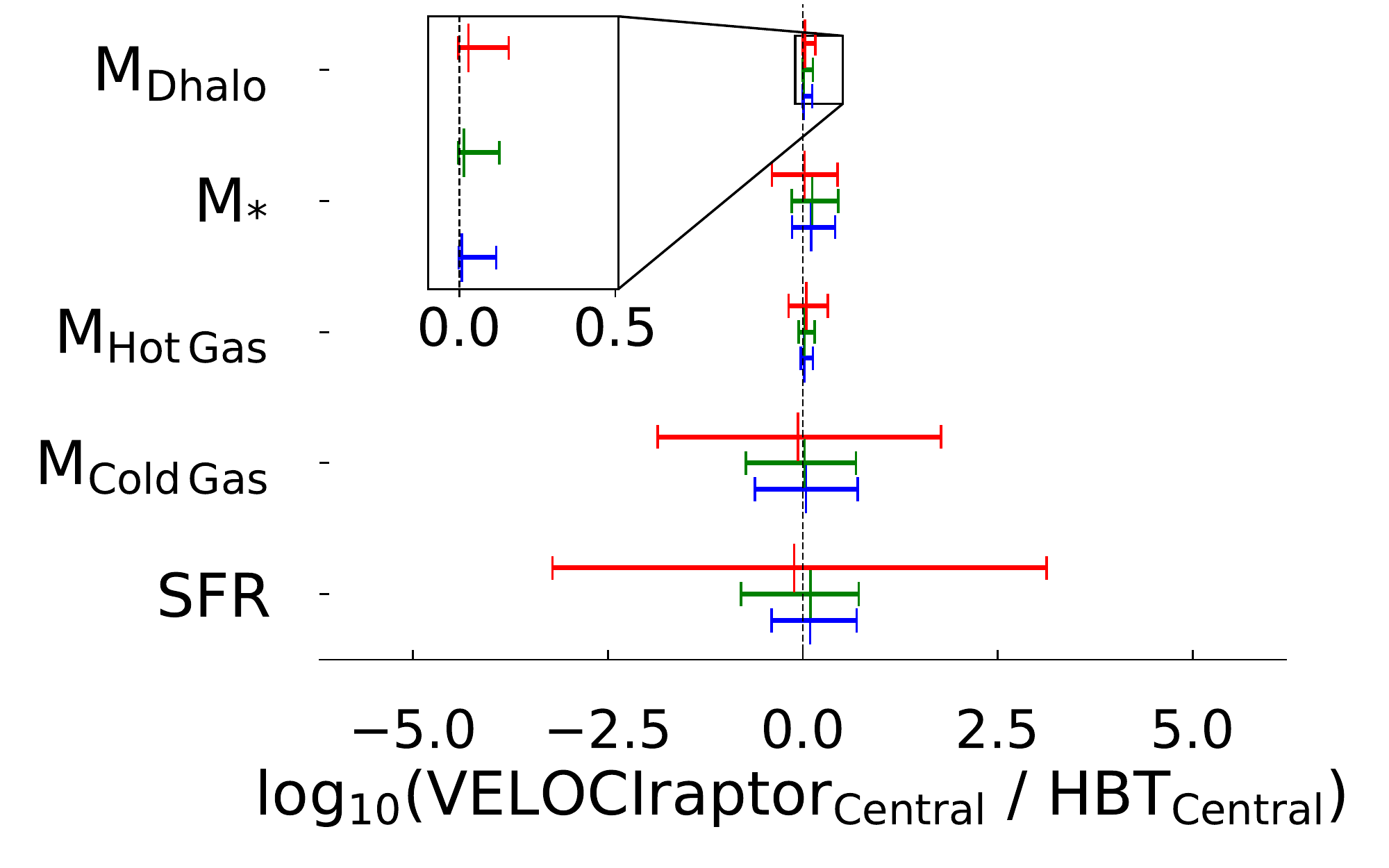}
    \includegraphics[scale=0.28]{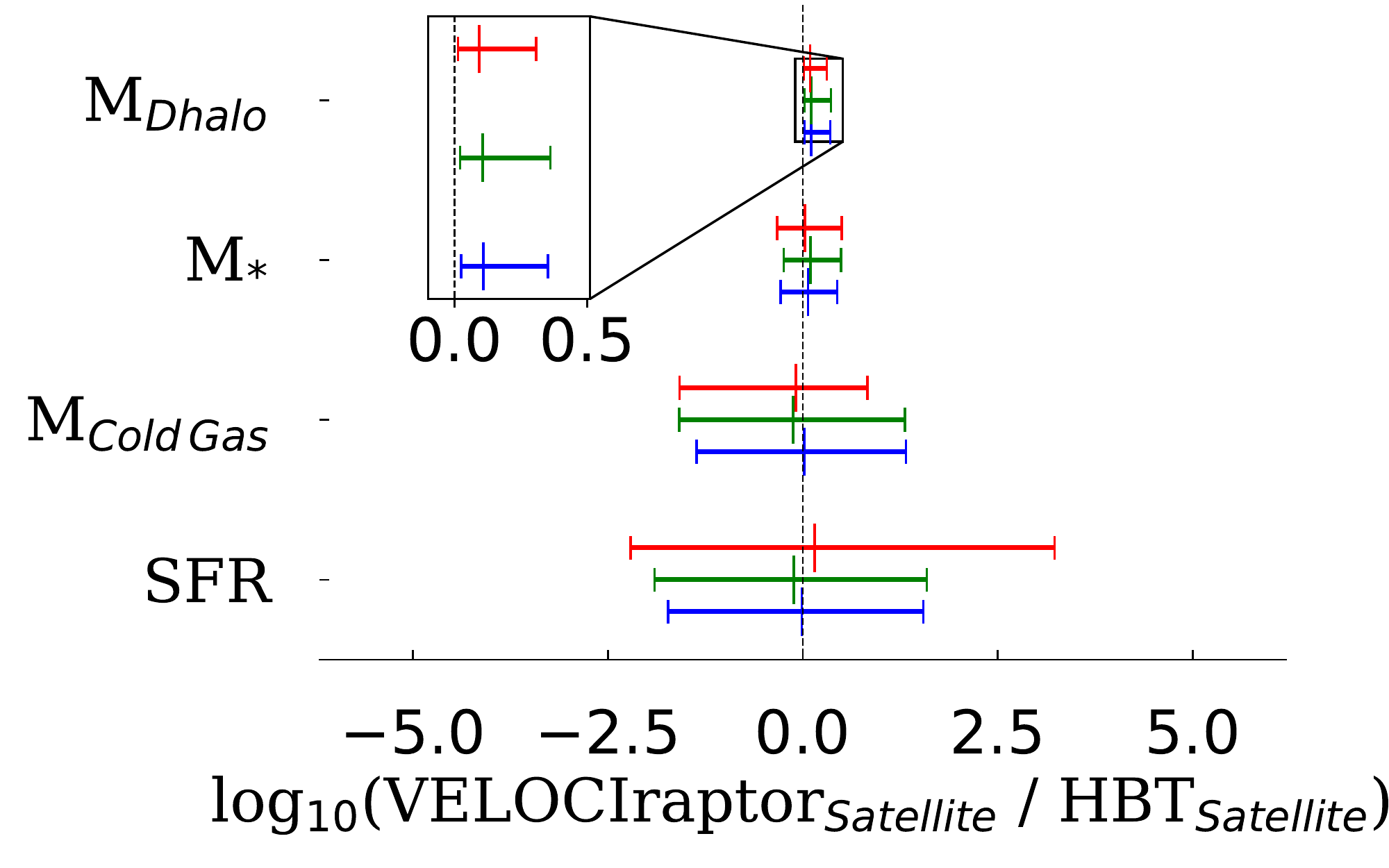}
    \includegraphics[scale=0.28]{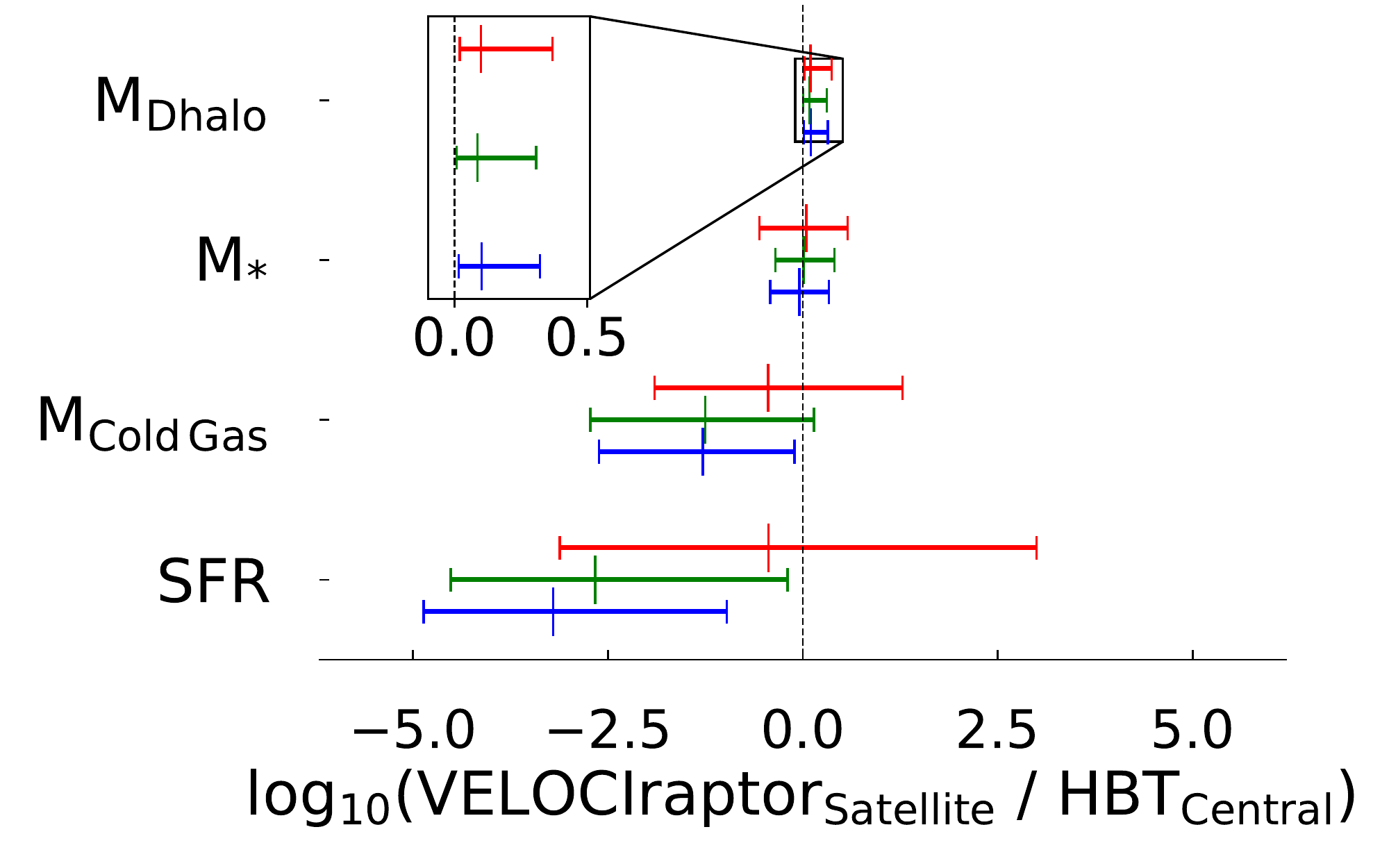}
    \caption{
    Ratios of Dhalo masses, galaxy stellar masses, hot gas masses (for central to central comparisons), and star formation rates, 
    for galaxies matched via their subhaloes between \hbt~ and \rockstar, \subfind~ and \velociraptor~ respectively at $z=0$ (top, middle and bottom rows). Results are shown for three different ranges of stellar mass (measured for \hbt), as indicated in the key. The errorbars show the central 80 percent of the population of matched objects. Left: centrals in \hbt~ {that are matched with centrals in} the other finders; middle: satellite \hbt~ galaxies matched exclusively with satellite galaxies; right: centrals in \hbt~ matched with satellite galaxies in \rockstar, \subfind~and \velociraptor.  Different colours show ranges in \hbt~ stellar mass as indicated in the key. Each panel also shows a close-up for ratios of Dhalo masses since they have values close to 1.
    }
    \label{Fig: percentages mixed}
    \end{figure*}

    \begin{figure*}
    \centering
    \includegraphics[scale=0.255]{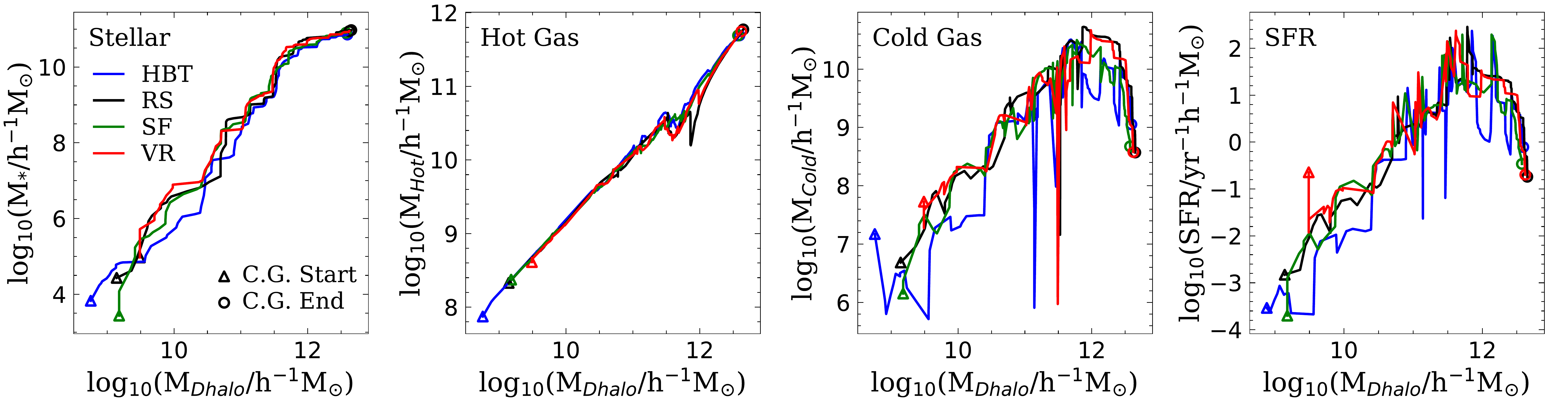}
    \includegraphics[scale=0.255]{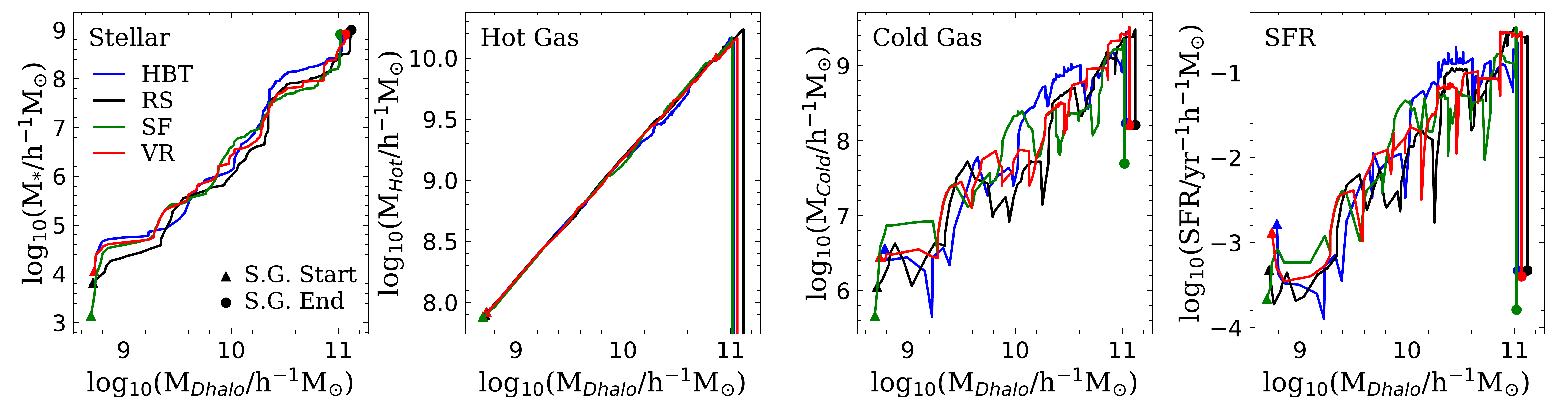}
    \includegraphics[scale=0.255]{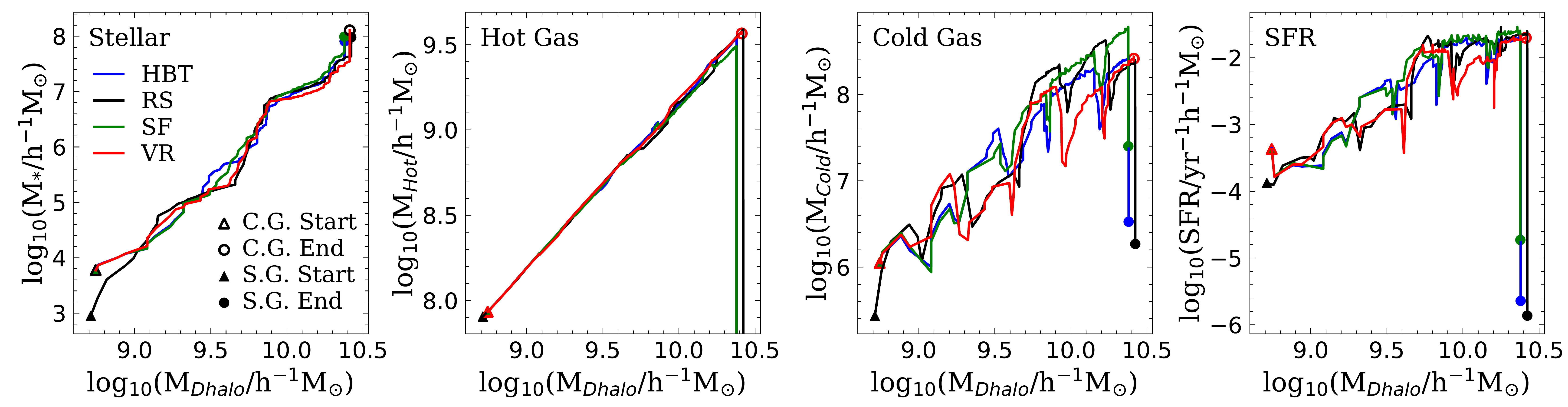}
    \caption{
    Evolution of Dhalo, stellar, hot gas and cold gas masses and SFRs   for galaxies matched between halo finders at $z=0$, as labelled. {The x-axis shows the host halo mass, which corresponds to the Dhalo mass for central galaxies, and the Dhalo mass at infall for satellites.} Top panels: central galaxies from \hbt, \rockstar, \subfind~ and \velociraptor. Middle panels: type 1 satellite galaxies from \hbt, \rockstar, \subfind~and \velociraptor. Bottom panels: central galaxies in \velociraptor~ matched to type 1 satellite {galaxies} in \rockstar, \subfind~and \hbt. The triangles and circles represent the start and end of the trajectories, respectively. Open triangles and circles indicate a start or end as {a central}, whereas filled symbols indicate a start or end as a type 1 satellite galaxy.
    }
    \label{panel: trajectories}
    \end{figure*}

Fig.~\ref{fig:satdelta} shows {a} comparison of properties of galaxies hosted by satellite subhaloes in both \hbt~ and \subfind~ in the top row, and the case when central galaxies in \hbt~ are matched to satellite galaxies in \subfind~ in the bottom row. When  galaxies are satellites in both \hbt~and \subfind, satellite Dhalo infall masses show almost no differences in their medians between these two finders, and galaxy properties show only very slight  differences in their median values. When a galaxy is central or satellite in both \hbt~and \subfind, the width of the 10-90 percentile range between \hbt~and \subfind~properties of matched objects increases going from subhalo mass to stellar mass, and {then to} cold gas {mass} and SFR. But as the average is always centered on a ratio of unity, the average properties {are} similar for the different finders, even for cold gas mass and SFR. When the galaxy is {a} central in \hbt~and {a} satellite {in} \subfind, then the Dhalo  masses ({at infall} in the case of the satellites) are also similar, but very slightly larger in \hbt. However, stellar {and} cold gas masses and SFRs are larger for \hbt~centrals than for the \subfind~satellites that they are matched with, particularly for lower mass galaxies, which is reasonable taking into account that \galform~removes hot gas mass from a galaxy as it becomes a satellite, which restricts the amount of star formation and hampers the growth of stellar mass for the satellite. The amplitude of the difference is higher in increasing order {of} Dhalo mass, stellar mass, cold gas mass and SFR.

This comparison can be made between all {halo} finders. A summary for central galaxies can be seen in the left column of Fig. \ref{Fig: percentages mixed}, which shows the variation of galaxy properties (\mstellar, \mhotgas~and \inststr) and Dhalo masses (M$_{\mathrm{Dhalo}}$) between matched central galaxies at $z=0$. 
The difference in stellar masses can in part explain the slight differences in the high mass tail of the stellar mass functions shown in Fig. \ref{fig: csmf}, where the drop in space density occurs at higher masses in \rockstar, followed by \velociraptor, \subfind~and then \hbt. It is clear that the variation in Dhalo mass that comes from the halo finder is  small in comparison to the scatter in galaxy properties between finders, which is probably due to the accumulated effects of variations across cosmic time between the merger trees based on the different finders. Hot gas masses also show a small scatter as they are closely related to the Dhalo mass. The scatter in each galaxy property is not sensitive to  the finder; as scatter can change the steep part of a distribution function, it is safe to say that differences in the stellar mass functions are not due to different scatters in stellar masses. We have also looked at the distributions of hot gas mass, cold gas mass and SFR for galaxies in different stellar mass ranges (not shown here) and they are consistent in shape between the halo finders, particularly in the high cold gas mass, hot gas mass, and SFR range which is the most sensitive to the scatter as it would widen the distribution to higher values. There are only small differences in the tails but these represent very small fractions of the galaxy population. 

For central galaxies matched to central galaxies, the average offset in galaxy properties between finders does not vary strongly with stellar mass. The scatter in stellar mass, cold gas mass and SFR between pairs of halo finders does tend to be larger for higher stellar masses. Recall, higher stellar masses imply  merger trees that {began to} form earlier and have longer branches; this increases the probability of finding larger integrated differences between the tree builder codes. {The stellar masses of central galaxies in \subfind~are very similar to those for \hbt, whereas for \rockstar~and \velociraptor~they are slightly higher than for \hbt. }

The middle column of Fig.~\ref{Fig: percentages mixed} shows the variation in galaxy properties for satellite galaxies in \hbt~ matched to satellites in the other halo finders. Here the comparison does not include hot gas because this is completely stripped off in \galform~ as subhaloes become satellites. The differences are again small for the Dhalo infall mass, and larger for the properties of galaxies. These differences are noticeably larger than for matched centrals, and are of increasing amplitude for stellar and cold gas masses, and largest for the SFR. The scatter is slightly larger for higher stellar masses. The stellar masses of satellite galaxies in \rockstar~and \velociraptor~are slightly higher than for \hbt~{as was found for central galaxies}.

Differences are also present when comparing \hbt~central galaxies matched with satellite galaxies in the other halo finders. This is shown in the right column of Fig.~\ref{Fig: percentages mixed}, where it can be seen that \hbt~ shows higher Dhalo, stellar, and cold gas masses and SFRs than \subfind.  This is reasonable given that central galaxies in \galform~ can continue to accrete baryonic matter due to gas cooling while satellite galaxies cannot. The differences are similar between \hbt~and two other finders, \rockstar~and \velociraptor, with larger masses for \hbt~as in its case the galaxies are centrals. The infall Dhalo masses are the properties that are matched best, and the scatter is much smaller than for the galaxy properties.  It should be noted that compared to matched centrals (left column), the galaxy properties of centrals matched to satellites show a much larger scatter.  As will be shown later, the cases where a central is matched to a satellite galaxy typically correspond to the time at which the galaxy is about to become a satellite; because of this, even though instantaneous properties such as cold gas mass and SFR show strong differences, the stellar mass which is an integrated property shows much smaller differences and a similar scatter as in the satellite to satellite and central to central comparison.  When looking at higher stellar masses $>10^{10}$\Mh, even the cold gas masses and SFRs become consistent, possibly because the cold gas mass fraction at these masses is much lower due to AGN feedback.

\subsection{Comparing evolution of individual galaxies}

 {Galaxies hosted by matched subhaloes behave in a similar way} between the finders. Fig.~\ref{panel: trajectories} shows the evolution of three different galaxies matched between the finders.  
The top panels show the trajectories of properties of a galaxy that is a central for all halo finders at $z=0$, with a final stellar mass $\sim 10^{10}$\Mh. The middle panels show the evolution of properties of a galaxy that is a satellite for all halo finders at $z=0$, at which point its stellar mass is $\sim 10^{9}$\Mh. The bottom panels show the evolution of properties of a galaxy that in \velociraptor~is a central galaxy at $z=0$, which is matched to satellite galaxies in the other finders.
This galaxy shows the lowest final stellar mass of the three examples, $\sim 10^{8}$\Mh.

The top panels of Fig.~\ref{panel: trajectories} show that the evolution of central galaxies is quite similar among the four finders; even though the subhalos were matched mostly by their positions, the final $z=0$ stellar masses are quite similar for all finders.  There are differences at earlier snapshots that can be of more than one order of magnitude for the stellar mass at fixed subhalo mass, but these differences are not present in the hot gas mass, which is always in excellent agreement between the different finders (hot gas mass is the most stable quantity under changes of finder as we saw in Fig \ref{Fig: percentages mixed} since it depends more directly on Dhalo mass). The cold gas mass is also reasonably similar among the different finders, although there are considerably larger fluctuations than for hot gas and stellar mass.  The cold gas mass reaches a maximum value around $M_{\rm Dhalo}\sim 10^{12}$\Mh~  regardless of the halo finder, which corresponds to approximately the point where the stellar masses reach a near plateau in the top-left panel,  due to the onset of AGN feedback which kicks in at a similar moment in all four finders. This maximum is also present in the SFR  with a subsequent drop to lower values.

A similar comparison is seen in the middle row of Fig.~\ref{panel: trajectories}, which shows an example of a galaxy that is a satellite in all halo finders, with  galaxy properties  showing no strong differences between halo finders; \mhotgas shows a drop to zero as the galaxies become satellites at $M_{\rm Dhalo}\sim 10^{11}$\Mh, a point where \mcoldgas~and the SFR show a decrease, since the removal of hot gas when a galaxy becomes a satellite results in star formation and SNe feedback depleting the cold gas reservoir, which inevitably produces a downturn in the star formation rate. Note that, by construction, the Dhalo mass at infall remains constant after infall, which is the reason why the Dhalo mass does not decrease in this plot \citep{Helly2003}. 

In the bottom panels of Fig.~\ref{panel: trajectories} the evolution of the galaxy with \velociraptor~is different to that with the other finders because it is a central only in \velociraptor, and therefore it has higher \mhotgas~ and \inststr~ at the last step. The differences are small except at the later steps when the galaxy becomes a satellite in the other three finders.  In most cases where a galaxy is a central in only one of the four finders it is because the time is quite close to when a subhalo changes from a central to a satellite, and the exact moment that this change takes place depends on the halo finder.

\section{Conclusions}
\label{conclusions}

We investigated the effect of using different algorithms to identify dark matter halos and construct merger trees on the trees themselves and on predictions for galaxy properties from the \galform\ semi-analytical model \citep{Cole2000}. The tree building algorithm can influence the output of \galform. {Before running \galform, the outputs from the different halo finding and tree building algorithms are processed through the Dhalo algorithm, which groups subhaloes into the top-level virialized halos called Dhaloes and builds merger trees for these. The processing of the halo trees also imposes the requirement that the mass of a Dhalo increases monotonically with time, and classifies the subhaloes within a Dhalo as being either a central or satellites. }
The processed trees are fed into \galform. We studied four merger tree builders, defined as a combination of halo finder + tree builder: 
\subfind, \velociraptor, \rockstar~and \hbt. 
%

Overall, despite applying different algorithms to identify subhaloes, their resulting properties show only slight differences. {The different} algorithms find {mostly} the same subhaloes because 
{the subhaloes are real dynamical structures.} Therefore, the differences in the properties between these subhaloes are due to the different options adopted by the different finders to assign, for instance, masses. For example, \rockstar~and \velociraptor~use a (6D) phase space search to identify subhalos after using the 3D FoF algorithm (see sections \ref{subsubsection: rockstar} and \ref{velociraptor: halo finder}). \subfind~and \hbt, on the other hand, use only 3D information to identify subhaloes.

Just as halo finders find subhaloes with slight differences in their properties, the choices adopted by each halo finder can cause certain subhaloes to be missed, since they do not meet the requirement to be selected. Phase-space halo finders are {in general} better able to find subhaloes in difficult conditions such as in high density regions within larger halos. The choices adopted by each halo finder not only affect the number of subhaloes found but also their classification as main or satellite subhaloes. Therefore, even though two different halo finders may find the same subhalo, it can have a different hierarchical classification, being classified as a satellite subhalo in one halo finder and as a main subhalo in the other.

The choices in each algorithm lead to only slight changes in the cumulative subhalo mass functions for main subhaloes, with \rockstar~ finding slightly more massive main subhaloes than the other finders. Bigger differences are found for satellite subhaloes, with \velociraptor~finding the most, particularly at high redshift.

Our analysis of the Dhalo processed outputs of the tree finders shows that at $z=0$ there are {only} slight differences in the distributions of Dhalo masses, {apart from} for \velociraptor~{which shows a} higher abundance of Dhaloes compared to the other finders. Objects that become satellite subhaloes are classed as type~1 if the subhalo is still identified by the halo finder, and as type~2 if it is no longer detected. In either case, \galform\ calculates galaxy properties based on the Dhalo mass of the satellite at infall.
At $z=0$ \hbt, \subfind\ and \rockstar\ have very similar type~1 satellite Dhalo mass functions, while \velociraptor\ results in many more  type~1 satellites than the other finders for Dhalo  masses above  $10^{13}$\Mh. The abundance of type~2 subhaloes as a function of infall Dhalo mass is significantly higher in \velociraptor, followed by \rockstar, with \hbt\ usually showing the lowest abundances. 

We then studied how the properties of the {Dhalo} merger trees 
affect the galaxy population predicted by \galform. 
Once a galaxy becomes a type 2 satellite, because its host subhalo can no longer be resolved, \galform~calculates how much longer it will survive before merging with the central galaxy using an analytical estimate of the dynamical friction timescale. This timescale depends on the ratio of the satellite subhalo mass to the Dhalo mass as well as the position and velocity of the subhalo at the last time at which it was identified; we looked at the satellite subhalo to Dhalo mass ratio for satellite subhaloes that are merging with central subhaloes, and find practically no differences between the finders.  

These findings point to the choice of halo finder having only a small impact on the galaxy population, after processing by \dhalo.
The results of \galform~show that the number of central galaxies does not  depend strongly on the halo finder or the definition of {main} and satellite subhalos. 
The number of type 1 and 2 satellite galaxies does show a stronger dependence on the tree builder. The number of \velociraptor~type 2 satellite galaxies is higher than the other 3 tree builders, in agreement with the excess of type~2 satellite subhaloes seen in  \velociraptor. 

Other properties of the galaxy population show only a slight dependence on the halo and tree finder, such as the relation between stellar and \dhalo~mass, and the $r$-band half-light radius as a function of stellar mass. The \velociraptor~ run displays the strongest differences, which, nevertheless, are still small.
Larger differences between the output with different finders are found for the star formation rate, especially for satellite galaxies where  at $z=6$ there is an order of magnitude lower star formation rate density for the \hbt\ run compared to the \velociraptor\ run.  Central galaxies account for most of the star formation rate density, and show only a factor of $\sim 2$ difference between \rockstar~and \hbt~(the lowest one) even at redshifts as high as $z=7$. The excess of satellites in \velociraptor\ is  accompanied by a higher star formation rate density and smaller galaxy sizes, but the amplitude of these differences is small, and insufficient to have an impact on observational comparisons.

We also looked at the HOD of galaxy samples selected by stellar mass and SFR with two different space densities, and found that the occupation of centrals and type 1 satellites is very similar among the merger tree builders, with only a slightly higher average number of type 2 satellites in \velociraptor. Given that the abundance of type 2 satellites is low compared to all galaxies, and that the large scale clustering of galaxies is dominated by the mass at which centrals reach an occupation of unity, we expect that the clustering of \galform~galaxies is not affected by the choice of halo finder, provided that the halo trees are prepocessed by the Dhalo algorithm.
    
We also match individual galaxies from the four different runs. 
%
When comparing centrals matched to centrals and satellites matched to satellites, their average properties agree between the outputs from the different finders. Even though the scatter is fairly small for Dhalo mass, hot gas mass, and stellar mass, but much larger for cold gas mass and SFR, we find that the distribution of baryon properties does not vary significantly between finders. 

We have thus shown that if we ensure that the output of different halo and merger tree finders is properly homogenised via the Dhalo algorithm (a component of {the \galform~modelling framework}), the  \galform~predictions for galaxy properties do not change significantly. 
Therefore, it is safe to apply \galform, one of the main SAMs available today, to the merger trees from \hbt, \velociraptor, \subfind~ and \rockstar, that different groups make available for dark-matter only simulations to make uniform comparisons and predictions for upcoming surveys.

\section*{Acknowledgements}
This project received funding from the European Union's Horizon 2020 Research and Innovation Staff Exchange Programme under the Marie Sk\l{}odowska-Curie grant agreement No 734374.
JSG and NP acknowledge support from CONICYT project Basal AFB-170002.
JSG also acknowledges funding from the CONICYT PFCHA/DOCTORADO BECAS CHILE/2019 21191147.
NP was supported by Fondecyt Regular 1191813.
CGL, CMB and JCH acknowledge support from STFC (ST/T000244/1). This work used the DiRAC@Durham facility managed by the Institute for Computational Cosmology on behalf of the STFC DiRAC HPC Facility (www.dirac.ac.uk). The equipment was funded by BEIS capital funding via STFC capital grants ST/K00042X/1, ST/P002293/1, ST/R002371/1 and ST/S002502/1, Durham University and STFC operations grant ST/R000832/1. DiRAC is part of the National e-Infrastructure. This work also used the Geryon computer at the Center for Astro-Engineering UC, part of the BASAL PFB-06, which received additional funding from QUIMAL 130008 and Fondequip AIC-57 for upgrades. The EAGLE simulation was performed on the DiRAC-2 facility at Durham, managed by the ICC, and the PRACE facility Curie based in France at TGCC, CEA, Bruy$\grave{e}$resle-Ch$\hat{a}$tel.

\section{Data Availability}
The data underlying this article will be shared on reasonable request to the corresponding author.



\bibliographystyle{mnras}
\bibliography{references}

\bsp	
\label{lastpage}
\end{document}